\documentclass[acmsmall,authorversion,nonacm,authordraft,review=false]{acmart}

\AtBeginDocument{%
  \providecommand\BibTeX{{%
    \normalfont B\kern-0.5em{\scshape i\kern-0.25em b}\kern-0.8em\TeX}}}

\setcopyright{acmcopyright}
\copyrightyear{2018}
\acmYear{2018}
\acmDOI{XXXXXXX.XXXXXXX}

\usepackage{graphicx}
\usepackage[export]{adjustbox}

\acmConference[Conference acronym 'XX]{Make sure to enter the correct
  conference title from your rights confirmation emai}{June 03--05,
  2018}{Woodstock, NY}
\acmBooktitle{Woodstock '18: ACM Symposium on Neural Gaze Detection,
 June 03--05, 2018, Woodstock, NY} 
\acmPrice{15.00}
\acmISBN{978-1-4503-XXXX-X/18/06}

\begin{document}

\title{Data Cards: Purposeful and Transparent Dataset Documentation for Responsible AI}

\author{Mahima Pushkarna, Andrew Zaldivar, Oddur Kjartansson}
\email{mahimap@google.com, andrewzaldivar@google.com, oddur@google.com}
\affiliation{
\institution{Google Research}
\country{Canada, United States, United Kingdom}
}


\begin{abstract}
As research and industry moves towards large-scale models capable of numerous downstream tasks, the complexity of understanding multi-modal datasets that give nuance to models rapidly increases. A clear and thorough understanding of a dataset's origins, development, intent, ethical considerations and evolution becomes a necessary step for the responsible and informed deployment of models, especially those in people-facing contexts and high-risk domains. However, the burden of this understanding often falls on the intelligibility, conciseness, and comprehensiveness of the documentation. It requires consistency and comparability across the documentation of all datasets involved, and as such documentation must be treated as a user-centric product in and of itself. In this paper, we propose Data Cards for fostering transparent, purposeful and human-centered documentation of datasets within the practical contexts of industry and research. Data Cards are structured summaries of essential facts about various aspects of ML datasets needed by stakeholders across a dataset's lifecycle for responsible AI development. These summaries provide explanations of processes and rationales that shape the data and consequently the models—such as upstream sources, data collection and annotation methods; training and evaluation methods, intended use; or decisions affecting model performance. We also present frameworks that ground Data Cards in real-world utility and human-centricity. Using two case studies, we report on desirable characteristics that support adoption across domains, organizational structures, and audience groups. Finally, we present lessons learned from deploying over 20 Data Cards.x
\end{abstract}

\ccsdesc[500]{Social and professional topics~User characteristics}
\ccsdesc[300]{General and reference~Evaluation}
\ccsdesc[500]{Software and its engineering~Software creation and management}
\ccsdesc[500]{Human-centered computing}

\keywords{data cards, dataset documentation, transparency, responsible AI, datasheets, model cards}


\setcopyright{none}
\begin{teaserfigure}
    \centering
    \includegraphics[width=0.65\linewidth, frame]{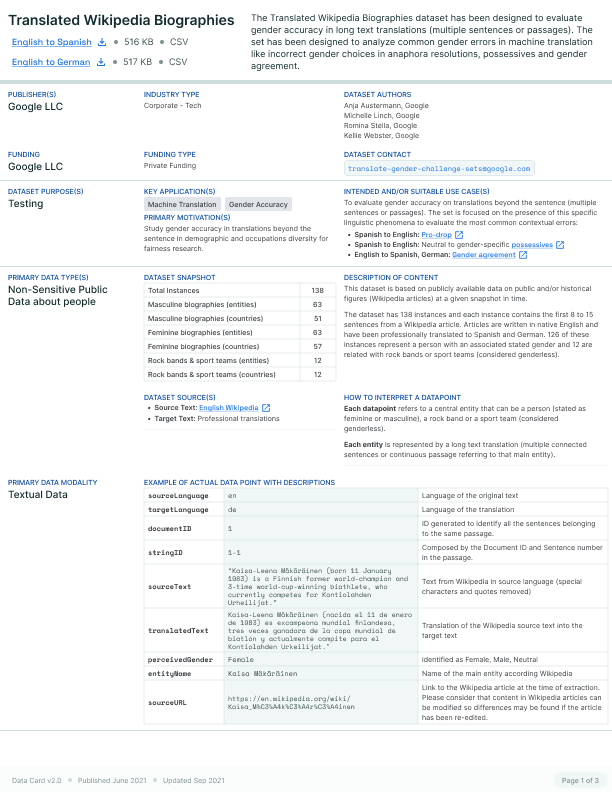}
    \caption{A page from a static, PDF data card describing a text dataset. The completed Data Card does not include questions from the template, but preserves its thematically contained row-and-column structure. Blocks increase in detail from left to right, and authors have introduced links to elegantly expose readers to additional documentation using context offered in the Data Card.}
    \label{fig:DataCard}
\end{teaserfigure}

\maketitle

\section{Introduction}

The challenge of transparency in machine learning (ML) models and datasets continues to receive increasing attention from academia and the industry \cite{facct, ainow}. Often, the goal has been to attain greater visibility into ML models and datasets by exposing source code \cite{antunes2018fairness}, contribution trails \cite{barclay2020framework}, introducing ML-drive data analysis methods \cite{kyd}, and introducing diverse oversight \cite{hutchinson2021towards}. 
Transparency and explainability of model outcomes through the lens of datasets has become a huge concern in regulation from government bodies internationally.  
However, attempts to introduce standardized, practical and sustainable mechanisms for transparency that can be create value at scale often meet limited success in research and production contexts. 
This reflects real world constraints of the diversity of goals, workflows, and backgrounds of individual stakeholders participating in the life cycles of datasets and artificial intelligence (AI) systems \cite{chander2018working, ehsan2021expanding, felzmann2020towards}. 

As a step towards creating value that connects dataset success to research and production experiences, we propose a new framework for transparent and purposeful documentation of datasets, called Data Cards. A Data Card contains a structured collection of summaries gathered over the life cycle of a dataset about observable (e.g., dataset attributes) and unobservable (e.g., intended use cases) aspects needed for decisions in organizational and practice-oriented contexts. Beyond metadata, Data Cards include explanations, rationales, and instructions pertaining to the provenance, representation, usage, and fairness-informed evaluations of datasets for ML models. These artifacts emphasize information and context that shape the data, but cannot be inferred from the dataset directly.  

Data Cards are designed as boundary objects \cite{star1989institutional} that should be easily discoverable, presented in an accessible format at important steps of a user journey for a diverse set of readers. They encourage informed decision making about data used when building and evaluating ML models for products, policy and research.  Data Cards complement other longer-form and domain-specific documentation frameworks for ethical reporting, such as Model Cards \cite{mitchell2019model}, Data Statements \cite{bender2018data}, and Datasheets for Datasets \cite{gebru2018datasheets}.

Data Cards are accompanied by frameworks that can be used to adapt them to a variety of datasets and organizational contexts. These frameworks are pivotal to establishing common ground across stakeholders to enable diverse input into decisions. Our case studies deomnstrate that creators of Data Cards were able to discover surprising future opportunities to improve their dataset design decisions, such as considering reasons for a high percentage of unknown values and the need to create a shared understanding of lexicons used in dataset labeling during problem framing within the team. 

In summary, our contributions are fourfold:
\begin{itemize}
\small
\item We document our multi-pronged development methodology in the setting of a large-scale technology company, and present a typology of stakeholders that span a typical dataset lifecycle. We translate these into corresponding objectives and principles for the creation of Data Cards to systematically reduce the knowledge asymmetries across stakeholders. 
\item We introduce a transparency artifact for at-scale production and research environments, \textbf{Data Cards}---structured summaries of essential facts about various aspects of ML datasets needed by stakeholders across a dataset’s lifecycle for responsible AI development, and describe the content (\textit{What information to present}), design (\textit{How to present information}), and evaluation (\textit{Assess the efficacy of information}) of Data Cards.
\item We propose three frameworks for the construction of Data Cards that focus on information organization, question framing, and answer evaluation respectively. Specifically, we describe OFTEn, our novel knowledge acquisition framework to arm dataset producers with a robust, deliberate, and repeatable approach for producing transparent documentation.
\item We present case studies on the creation of Data Cards for a computer vision dataset and a language dataset to demonstrate the impact of Data Cards as boundary objects in practice, and discuss the epistemic and organizational lessons learned in scaling Data Cards.
\end{itemize}

Our collective efforts suggest that in addition to comprehensive transparency artifacts\footnote{For the purposes of practicality, we use transparency artifacts as a general term to describe both Data and Model Cards because of their inextricably linked nature. In this paper, we primarily focus on our insights and advances on datasets and correspondingly Data Cards, our novel contribution.}, the creation of structured frameworks are not only beneficial in adding nuance to the dataset documentation process itself, but also transformational in introducing human-centric and responsible practices when using datasets in ML applications. 

\section{Development Methodology}
Over the course of 24 months, multiple efforts were employed to design Data Cards and its supporting frameworks, borrowing from methods in human-centered design, participatory design, and human-computer interaction. We worked with dataset and ML teams in a large technology company to iteratively create Data Cards, refining our design decisions to respond to challenges in production contexts. In parallel, we ran studies and workshops to identify opportunities and challenges in the implementation of Data Cards. In this section, we detail the various efforts and describe their impact on the development of Data Cards. 

Specifically, we worked with 12 teams in a large technology company to create 22 Data Cards that describe image, language, tabular, video, audio, and relational datasets in production settings. Teams ranged in size from four to over 20 members, and were comprised of some combination of research software engineers, research scientists, data analysts and data program managers. This allowed us to observe each teams' documentation workflows, collaborative information gathering practices, information requests from downstream stakeholders, review and assessment practices. Our co-creative approach in conjunction with feedback received across other studies yielded continuous improvements in the usability and utility of each new Data Card created.

As we worked with ML dataset and model owners to produce prototypical transparency artifacts, their drafts were evaluated in an external focus group with nine participants. These participants represented non-expert, technical use cases from User Experience (UX) and Human-Computer Interaction (HCI) research, Policy, Product Design \& Development, Academia, and Law. Participants were asked to complete a paper-based questionnaire to reflect on their ideals of transparency. This was used as a basis for a broader discussion on transparency. Participants were then provided with printed artifacts which they annotated with their feedback. This allowed us to capture specific feedback and establish relationships across themes and topics in the artifacts. We concluded with a discussion reflecting on their use of transparency artifacts and an offline survey to capture their overall expectations. Through this focus group, we were able to arrive at a working definition and values of transparency relevant to domains within AI product life cycles. We further synthesized feedback on the transparency artifacts into an initial set of recommendations to combat common reader-side challenges, which were then offered as guidance to teams creating Data Cards.

Based on our experience in co-creating Data Cards with teams, we were able to consolidate recurring and overlapping questions into a canonical template that documents 31 different aspects of data sets. Questions that are were modality-specific were consolidated into appendable blocks, but largely left out of the canonical template. A follow-up  MaxDiff survey (n=191) was conducted to understand the information needs in dataset documentation within our company. Through this survey, we learned the relative importance of the 31 different themes documented in a Data Card, how these vary by dataset modality and job function, further incorporated into our design of Data Cards. We observed the need for a generative framework that Data Card creators could use to add or tailor question to new datasets without compromising the readability, navigability, comparability and transparency intrinsic to the Data Card.

Our internal study recruited 30 experts spanning sixteen teams within our company. Participants represented stakeholders who (a) create datasets designed for ML use cases and (b) use or review datasets for applied and foundational model development. Over the course of three days, this group engaged in various participatory activities to capture their use cases for transparency artifacts, information requirements and strategies for evaluation of transparency artifacts. Participants were then invited to actively contribute to future discussions of Data Cards and their development as it related to the participant’s specific data domains. We found that despite their deep expertise and experience, participants were unable to provide examples of exemplary documentation, but were quick to furnish 'excellent' examples of poor documentation. This pointed us to the need for a set of dimensions that can be used to assess transparency and documentation without conflating documentation with the dataset.

Further, we developed a structured participatory workshop-based approach to engage cross-functional stakeholders when creating transparent metadata schema for dataset documentation.
This methodology was open-sourced and tested in the data domains of human computation, geo-spatial ML, multi-modal data operations, healthcare data, community-engaged research, and large-scale multitask language models. Common to all workshops, we found that teams adopting Data Cards often started with an intuition about the benefits of transparency in dataset documentation. To define the content, infrastructure, and processes for data card creation (and other complementary documentation artifacts) required that teams aligned on a shared definition of transparency, audience, and the audience’s requirements. We observed organization-specific factors that can impact long-term sustainability of scaling Data Cards, such as knowledge asymmetries between stakeholders, organizational processes that incentivize the creation and maintenance of documentation, infrastructure compatibility and readiness, and communication culture across and within stakeholder groups. While a detailed discussion of our participatory methodology to developing transparency metadata schemas \cite{craftplaybook} and survey is beyond the scope of this paper, we introduce relevant critical frameworks from our methodology. 

\subsection{Framing Transparency in the Context of Data Cards}
Despite the diverse backgrounds of participants across studies, a shared dominant perception was that transparency artifacts were, ironically, opaque. The opacity in documentation, quite simply, increases when language used is technical, dense, and presumptive of a reader’s background, making it difficult for non-technical stakeholders to interpret. This leads to sub-optimal decision making, and propagates asymmetries in power structures and myopic AI data practices. Participants described transparency as subjective, audience-specific and contextual. To that end, we frame our definition of transparency as \textit{“a clear, easily understandable, and plain language explanation of what something is, what it does and why it does that”}, to emphasize the domain-agnostic and inclusive prerogative of transparency artifacts. We present eight characteristics of transparency that are vital for a robust discussion of the benefits, values, ethics, and limitations of AI datasets (Table \ref{tab:transparency}).
\begin{table}
 \scriptsize
  \centering
  \caption{Characteristics of Transparency}
  \label{tab:transparency}
  \begin{tabular}{p{4cm}p{8cm}}
    \toprule
   Transparency  Characteristic&Description\\
    \midrule
    \ Balance opposites & For example, disclosing information about AI systems without leaving creators vulnerable beyond reason, reporting fairness analyses without legitimizing inequitable or unfair systems, introducing standards for transparency that are wholly automated or become checklists.\\
    \ Increase in expectations & Any information included in a transparency artifact can be expected to receive greater scrutiny.\\
    \ Availability and comfort & Users want access to transparency information at multiple levels, even if they don’t need to use it. \\
    \ Requires checks and balances & Transparency artifacts and their creation must be amenable to 3rd party evaluation, with the caveat that excessive transparency can open an AI system vulnerable to adversarial actors.\\
    \ Subjective interpretations & Stakeholders have different definitions and unique ideas on what constitutes transparency. \\
    \ Trust enabler & Accessible and relevant information about AI systems in-creases the the willingness of a data consumer or user to take a risk based on the expectation of benefits from the data, algorithms and the products they use. \\
    \ Reduce knowledge asymmetries & Cross-disciplinary stakeholders are more effective when they possess a shared mental model and vocabulary to describe aspects of the AI system. \\
    \ Reflects human values & It comes from both technical and non-technical disclosure about assumptions, facts and alternatives. \\
  \bottomrule
\end{tabular}
\end{table}
Data Cards aim to provide a singular framework that allows non-traditional stakeholders across product, policy, and research to understand aspects about datasets and how they are used to make informed decisions. 
We found that stakeholders review role-related topics in Data Cards with amplified scrutiny, and follow-up questions progressively increase in specificity.
This highlighted a fundamental principle underpinning transparency—that transparency is attained when we establish a shared and Socratic understanding of datasets based on the ability to ask and answer questions \textit{over time}. 

\subsection{A Typology of Stakeholders}
At first, our audience for Data Cards was fairly broad, comprising a mix of experts and non-experts.
Frameworks proposed by Suresh, et al \cite{suresh2021beyond} have distinguished higher-level domain goals and objectives from lower-level interpretability tasks, but are limited by their epistemological framing and vast scope.
We created a broad yet decomposable typology describing three stakeholders groups in a dataset's life cycle, allowing us to consider how cross-functional stakeholders engage in decision-making on the basis of a single transparency artifact. 

\textit{\textbf{Producers}} are upstream creators of dataset and documentation, responsible for dataset collection, ownership, launch and maintenance. Producers often subscribed to a single, informal notion of “users” of Data Cards—loosely characterized by high data domain expertise, familiarity with similar datasets, and deep technical knowledge. However, in practice, we find that only a few readers or \textit{\textbf{Agents}} actually meet all these requirements. 

\textit{\textbf{Agents}} are stakeholders who read transparency reports, and possess the agency to use or determine how themselves or others might use the described datasets or AI systems. After testing prototypes and proof of concepts with different audience groups, it became clear that agents with operational and reviewer needs were distinct categories, and includes stakeholders who may never directly use the dataset, but will engage with the Data Card (for e.g. reviewers or non-technical subject matter experts). Agents may or may not possess the technical expertise to navigate information presented in typical dataset documentation, but often have access to expertise as required.

Additionally, agents are distinct from \textit{\textbf{Users}}, who are individuals and representatives who interact with products that rely on models trained on dataset. They may consent to providing their data as a part of the product experience. Users require a significantly different set of explanations and controls grounded within product experiences. We therefore restrict the use of Data Card for agents with access to technical expertise, and encourage the use of alternative transparency artifacts for users that are designed exclusively for that purpose.

We further dis-aggregate these high-level groups to generate awareness and emphasize the unique decisions that each sub-group must make (Fig[\ref{typologytable}]). However, these groupings exist on a continuum and stakeholders may fall into more than one group concurrently, depending on the context. We used this typology to unearth assumptions that are often made about the rich intersectional attributes of individual stakeholders, such as expertise (e.g. novice or expert), data fluency (e.g. none to high), job roles (e.g. Data Scientist, Policy Maker), function performed vis-à-vis the data (Data Contributor, Rater), and goals or tasks (Publishing a dataset, Comparing datasets) when conceptualizing Data Cards. Usability studies across these groups revealed guidelines for the successful and appropriate adoption of Data Cards in practice and at scale. These are distilled into the following objectives for Data Cards: 
    \subsubsection{\textbf{O1. Consistent:}} Data Cards must be comparable to one another, regardless of modality or domain such that claims are easy to interpret and validate within context of use. A Data Card creation effort should solicit equitable information from all datasets. While deploying one-time Data Cards is relatively easy, we find that organizations need to preserve comparability when scaling adoption.
    \subsubsection{\textbf{O2. Comprehensive: }}Rather than being created as a last step in a dataset’s lifecycle, it should be easy to create a Data Card concurrently with the dataset. Further, the responsibility of filling out fields in a Data Card should be distributed and assigned to the most appropriate individual. This requires standardized methods that extend beyond the Data Card, to the various reports generated in the dataset’s lifecycle.
    \subsubsection{\textbf{O3. Intelligible and Concise: }}Readers have varying levels of proficiency\footnote{Proficiency is a combination of data fluency and domain expertise. Data fluency is described as the familiarity and comfort that readers have in working with data that is both, in or outside of their domain of expertise. The greater the comfort with understanding, manipulating, and using data, the greater the fluency. Domain expertise is defined as “knowledge and understanding of the essential aspects of a specific field of inquiry” \cite{mccue2014data} in reference to the domain of the dataset.} which affects their interpretation of the Data Card. In scenarios where stakeholders differences in proficiency, individuals with the strongest mental model of the dataset become de-facto decision makers. Tasks that are more urgent or challenging can reduce the participation of non-traditional stakeholders (See \ref{typologytable}) in the decision, which is left to “the expert”.  This risks omitting critical perspectives that reflect the situated needs of downstream and lateral stakeholders. A Data Card should efficiently communicate to the reader with the least proficiency, while enabling readers with greater proficiency to find more information as needed. The content and design of a Data Card should advance a reader’s deliberation process without overwhelming them, and encourage stakeholders cooperation towards a shared mental model of the dataset for decision-making.
    \subsubsection{\textbf{O4. Explainability, Uncertainty: }}Study participants reported that ‘known unknowns’ were as important as known facets of the dataset in decision making. Communicating uncertainty along with meaningful metadata was considered a feature and not a bug, allowing readers to answer questions such as \textit{“Is a specific analysis irrelevant to the dataset or were the results insignificant?”} or \textit{“Is information withheld because it is proprietary or is it unknown?}”. Clear descriptions and justifications for uncertainty can lead to additional measures to mitigate risks, leading to opportunities for fairer and equitable models. This builds greater trust in the dataset and subsequently, its publishers \cite{bhatt2021uncertainty}.

\section{Data Cards}
Data Cards capture critical information about a dataset across its life cycle. Just as is true with every dataset, each Data Card is unique, and no single template satisfactorily captures the nuance of all datasets. In this section, we introduce our guiding principles, and offer a three-pronged description that elaborates on the design, content, and evaluation of Data Cards. We introduce corresponding frameworks that allow Data Cards to be tailored but preserve the utility and intent of Data Cards.
\subsection{Principles}
While most previous approaches take domain-specific \cite{bender2018data} or approaches \cite{gebru2018datasheets} that have been prescriptively adopted as a way of creating transparency artifacts, our novel contributions are the generative design of Data Cards as an underlying framework for transparency reporting for readability and scaling in production contexts. To meet the objectives stated above, Data Cards have been designed along the following principles:
\begin{itemize}
\item \textbf{P1. Flexible: }Describe a wide range of datasets – live or static, datasets that are actively being curated from single or multiple sources, or those with multiple modalities.
\item  \textbf{P2. Modular: }Organize documentation into meaningful sections that are self-contained repeatable units, able to provide an end-to-end description of a single aspect of the dataset.
\item \textbf{P3. Extensible: }Components that can be easily reconfigured or extended systematically for novel datasets, analyses, and platforms. 
\item  \textbf{P4. Accessible: }Represent content at multiple granularities so readers can efficiently find and effectively navigate detailed descriptions of the dataset.
\item  \textbf{P5. Content-agnostic:} Support diverse media including multiple choice selections, long-form inputs, text, visualizations, images, code blocks, tables, and other interactive elements. 
\end{itemize}

\subsection{Design and Structure}
\begin{figure}
  \includegraphics[width=\textwidth]{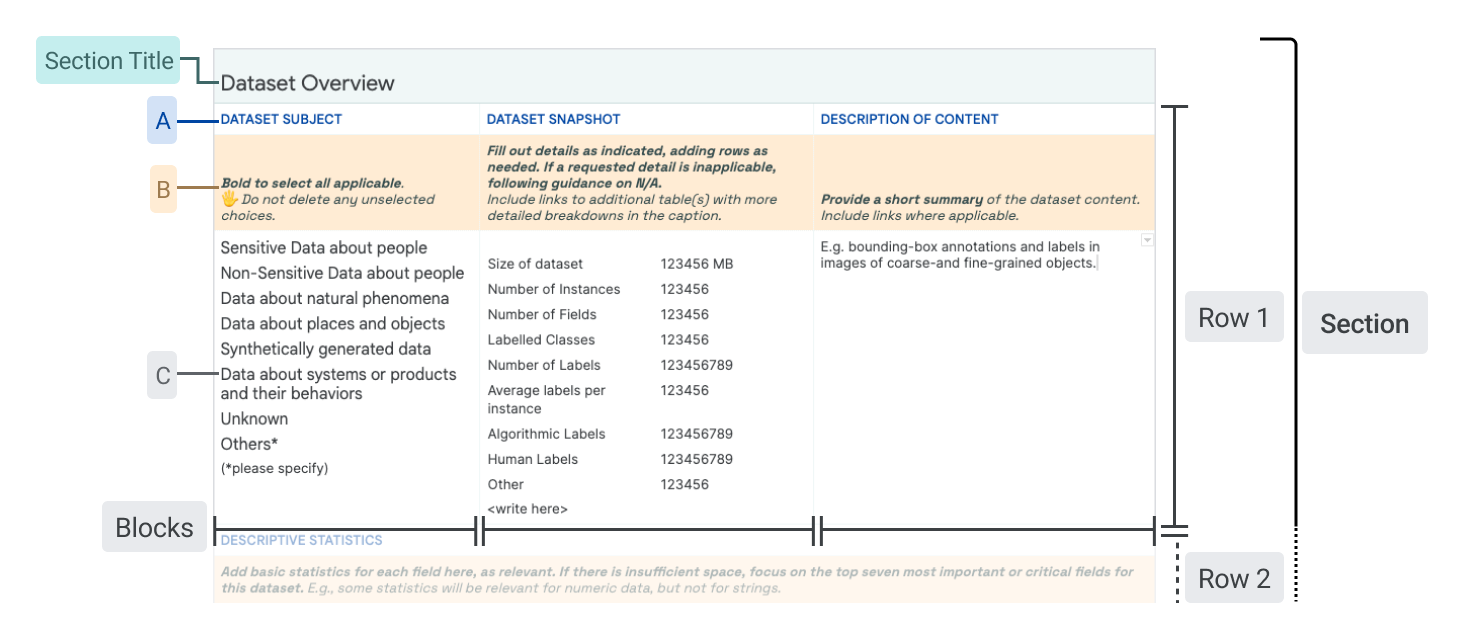}
  \caption{\textbf{A Data Card Template Section:} This section is titled "Dataset Overview", and contains two rows. The first row has three blocks, whereas the second row spans the entire width of the section. Blocks contain (A) A Title, (B) A prompting question, and (C) an answer input space populated with predetermined choices or suggested answer structures.}
  \Description{A sample section of an unfilled Data Card, with questions and answer prompts to describe the Dataset Overview.}
  \label{fig:teaser}
\end{figure}

The fundamental "display" unit of a Data Card is a \textit{\textbf{block}} which consists of a title, a question, space for additional instructions or descriptions, and an input space for answers. 
Answer inputs are reinforced with structure to create blocks that are specifically suited for long- or short-form text, multiple or single choice responses, tables, numbers, key value pairs, code blocks, data visualizations, tags, links, and demos of the data itself, in alignment with principles (\textbf{P2}) and (\textbf{P5}).  
In our templates, we iteratively introduced structures for open-ended answers, predetermined responses for multiple choice questions, and demonstrative examples where responses could be complex (Fig. \ref{fig:teaser}). Producers found these assistive efforts as useful guides for setting expectations about consistency, clarity, and granularity in responses. 
When completed, blocks typically retained titles and answers (See fig \ref{fig:DataCard}).

Blocks are arranged thematically and hierarchically on a grid to enable an “\textit{overview first, zoom-and-filter, details-on-demand}” \cite{shneiderman2003eyes} presentation of the dataset. 
In our template, blocks with related questions are organized into \textit{rows}, and rows are stacked to create \textit{sections} using meaningful and descriptive titles (Figure \ref{fig:teaser}). 
Each row is thematically self-contained so readers can effectively navigate multiple facets of a dataset in a Data Card. 
Answers increase in both detail and specificity across columns in the direction of the language in which the Data Card is written, allowing readers to find information at the appropriate fidelity for their tasks and decisions. Where appropriate, a single block may span multiple columns.
Sections are vertically arranged based on functional importance in a nested hierarchy marked by section titles in the first Data Card [\ref{DC1}]. 
Here, all necessary sections (dataset snapshot, motivations, extended use, collection and labeling methods) are established in order to provide greater context for interpreting sections that describe fairness-related analyses (fairness indicators, bounding box sizes). 
In contrast, sections in the second Data Card [\ref{DC2}] are organized in a flat hierarchy, suggesting equal importance of all blocks.
Variation within the formatting of the content communicates both denotative and connotative meaning. 

\subsubsection{Socratic Question-Asking Framework: Scopes}\label{scopes-framework}
To maintain Data Cards' accessibility \textbf{\textit{(P4)}} and enable agents with varying proficiency levels to progressively explore content, new information needs to be introduced at multiple levels of abstraction.
Further, adding ad-hoc blocks risks structurally compromising Data Cards for readers and producers alike, and can reduce its usability and integrity. 
Pertinent to objectives \textbf{\textit{O2}} and \textbf{\textit{O3}}, we provide a structured approach to framing and organizing questions to address common challenges in adapting Data Card templates for new datasets. 
Depending on its specificity, a new theme is deconstructed or a single question extrapolated into at least three questions at varying granularities, which we characterize as telescopes, periscopes, and microscopes.  
For the purposes of demonstration, we consider the documentation of sensitive human attributes:

\textbf{Telescopes} provide an overview of the dataset. These are questions about universal attributes that are applicable across multiple datasets, for example "\textit{Does this dataset contain Sensitive Human Attributes?}". Telescopes can be framed as binary (\textit{contains, does not contain}) or multiple choice (\textit{Select all that apply: Race, Gender, Ethnicity, Socio-economic status, Geography, Language, Sexual Orientation, Religion, Age, Culture, Disability, Experience or Seniority, Others (please specify)}). These serve three specific purposes. 
First, telescopic questions generate enumerations or tags that are useful for knowledge management, indexing and filtering in large repository of Data Cards. Second, they introduce and set context for additional information within a row, helping readers navigate larger or more complex Data Cards. Lastly, telescopic questions introduce conditional logic to streamline the experience of filling out a Data Card. When viewed together, telescopic questions offer a shallow but wide overview of the dataset.

\textbf{Periscopes} provide greater technical detail pertaining to the dataset. These are questions about attributes specific to the dataset that add nuance to telescopes. For example,\textit{ “For each human attribute selected, specify if this information was collected intentionally as a part of the dataset creation process, or unintentionally not explicitly collected as a part of the dataset creation process but can be inferred using additional methods)”}. A periscopic question can ask for operational information such as the dataset’s shape and size, or functional information such as sources or intentions. Responses typically look like key-value pairs, short descriptions, tables, and visualizations. Since periscopes often describe analysis results, statistical summaries, and operational metadata, they are often reproducible and can be automated wherein automating generates results that are more accurate or precise than human input.

\textbf{Microscopes} offer fine-grained details. These are questions about the “unobservable” human processes, decisions, assumptions and policies that shape the dataset. These elicit detailed explanations of decisions or summarize longer process documents that governed responses to the corresponding periscopic questions. For example, “\textit{Briefly describe the motivation, rationale, considerations or approaches that caused this dataset to include the indicated human attributes. Summarize why or how this might affect the use of the dataset.}”. Necessarily, these are difficult to automate unless there are standardized terms or operating procedures. Answers to microscopes are typically long-form text with lists and links, data tables, and visualizations. 

We find that the interpretations of a Data Card are greatly influenced by the presence or absence of these levels of abstraction. Telescopic questions are easiest to answer, but offer relatively low utility. Periscopic questions facilitate quick assessments of suitability and relevance of the dataset, essential for simple decision-making. We observed that microscopic questions were most challenging to answer since they require articulating implicit knowledge. Yet, these questions enabled agents and producers alike to assess risk, plan mitigations, and where relevant, identify opportunities for better dataset creation. Together, telescopes, periscopes, and microscopes layer useful details such that numerous readers can navigate without losing sight of the bigger picture. 

\subsection{Content and Schema}

Our initial approach was to create a single template to capture the provenance, intentions, essential facts, explanations and caveats in an accessible and understandable way. In co-creating Data Cards for different types of datasets, we identified 31 broad, generalizable themes (Table \ref{tab:content_schema}) that can comprehensively described any dataset (\textbf{O2}). However, these vary in importance on a per-task basis to stakeholders. Sections in our template (\ref{DC-template}) demonstrate how these themes are broken down into sets of scopes (\ref{scopes-framework}). To illustrate the differences in descriptions of the same theme in Data Cards, we include Data Cards from our case studies (\ref{cs1}, \ref{cs2}) in our appendix (\ref{DC1}, \ref{DC2}). 

\renewcommand{\arraystretch}{1.3}
\begin{table}[!ht]
    \centering
    \scriptsize
    \caption{Content themes in the Data Card template}
    \label{tab:content_schema}
    \begin{tabular}{p{6cm}p{6cm}}
    \toprule
    (1) About the publishers of the dataset and access to them & (17) The data collection process (inclusion, exclusion, filtering criteria)\\
    (2) The funding of the dataset & (18) How the data cleaned, parsed, and processed (sampling, filtering, etc.)\\
    (3) The access restrictions and policies of the dataset & (19) How the data was rated in the dataset, its process, description and/or impact\\
    (4) The wipeout and retention policies of the dataset & (20) How the data was labelled in the dataset, its process, description and/or impact\\
    (5) The updates, versions, refreshes, additions to the data of the dataset & (21) How the data was validated in the dataset, its process, description and/or impact\\
    (6) Detailed breakdowns of features of the dataset & (22) The past usage and associated performance of the dataset (eg. models trained)\\
    (7) If there are attributes missing from the dataset or the dataset’s documentation & (23) Adjudication policies related to the dataset (labeller instructions, inter-rater policies, etc.\\
    (8) The original upstream sources of the data & (24) Regulatory or compliance policies associated with the dataset (GDPR, licensing, etc.)\\
    (9) The nature (data modality, domain, format, etc.) of the dataset & (25) Dataset Infrastructure and/or pipeline implementation\\
    (10) What typical and outlier examples in the dataset look like & (26) The descriptive statistics of the dataset (mean, standard deviations, etc.)\\
    (11) Explanations and motivations for creating the dataset & (27) Any known patterns (correlations, biases, skews) within the dataset\\
   (12) The intended applications of the dataset & (28) Any socio-cultural, geopolitical, or economic representation of people in the dataset\\
    (13) The safety of using the dataset in practice (risks, limitations, and trade-offs) & (29) Fairness-related evaluations and considerations of the dataset\\
    (14) The maintenance status and version of the dataset & (30) Definitions and explanations for technical terms used in the dataset’s documentation (metrics, industry-specific terms, acronyms)\\
    (15) Difference across previous and current versions of the dataset & (31) Domain-specific knowledge required to use the dataset\\
    (16) Expectations around using the dataset with other datasets or tables (feature engineering, joining, etc.) & \\
    \bottomrule
    \end{tabular}
\end{table}

\subsubsection{OFTEn Framework}
Over time, we found it necessary to develop a consistent and repeatable approach to identify and add new themes from dataset life cycles in a Data Card that are reportable by everyone in the organization. Additionally, certain topics, such as consent, can span entire dataset life cycles with different implications at each stage. We introduce OFTEn, a conceptual tool for logically considering how a topic can promulgate across all parts of a Data Cards (\textbf{P1, P3}), that can be used inductively and deductively for detailed dataset transparency investigations. 

OFTEn (Table \ref{table: often-description}) is an acronym to describe common stages in the dataset life cycle ("Origins, Factuals, Transformations, Experience, and n=1 \textit{example}"). Though ordered, stages are loosely defined to mirror typical non-linear dataset development practices. Of note, how agents use the dataset practice is considered a distinct stage in OFTEn, affording the flexibility to incorporate feedback from downstream stakeholders (dataset consumers, product users, and even data contributors). This establishes a trail to track the performance of AI systems trained and evaluated on the dataset, and exposes any caveats or limitations that agents should be aware of.  

An OFTEn analysis of the dataset can preemptively enable the discovery of insights that would otherwise not be generally evident. Inductively, OFTEn supports activities with agents to formulate questions about datasets and related models that are important for decision-making. At its simplest, it can be visualized as a matrix in which rows represent the dataset life cycle, and columns provide prompts to frame questions (who, what, when, where, why, and how) about a given topic in the dataset’s lifecycle (Table \ref{table: often-description}. Its participatory use enables reporting both dataset attributes and implicit information that can affect outcomes in real-world deployment. Deductively, we use OFTEn to assess if a Data Card accurately represents the dataset, resulting in formative effects on both, documentation and dataset. Lastly, we find that Data Cards with a clear underlying OFTEn structure are easy to expand and update. This structure allows Data Cards to capture information over time, such as feedback from downstream agents, notable differences across versions, and ad-hoc audits or investigations from producers or agents.

\begin{table}[!ht]
\centering
\scriptsize
\caption{The OFTEn framework}
\label{table: often-description}
\begin{tabular} {p{0.2\textwidth}p{0.4\textwidth}p{0.3\textwidth}}
\hline
 & Description & Themes \\
 \hline
 \textbf{Origins} & Various planning activities such as defining requirements, design decisions, collection or sourcing methods, and deciding policies which dictate dataset outcome & \textit{Authorship, Motivations, Intended Applications, Unacceptable uses, Licenses, Versions, Sources, Collection Methods, Errata, Accountable parties} \\
  \hline
 \textbf{Factuals} & Statistical and other factual attributes that describe the dataset, deviations from the original plan, and any pre-wrangling analysis & \textit{Number of Instances, Number of Features, Number of Labels, Breakdown of subgroups, Description of features, Taxonomies of labels, Missing/Duplicates, Inclusion and exclusion criteria} \\
  \hline
 \textbf{Transformations} & Various operations such as filtering, validating, parsing, formatting, and cleaning through which raw data is transformed into a usable form including labeling or annotation policies, validation tasks, feature engineering and related modifications & \textit{Rating or Annotation, Filtering, Processing, Validation, Synthetic features, Handling of PII, Sensitive Variables, Fairness Analyses, Impact Assessments, Skews \& Biases} \\
  \hline
 \textbf{Experience} & Dataset is benchmarked or deployed in experimental, production, or research practice, including specific tasks, access training requirements, modifications made to suit the task, analyses, unexpected behaviors, limitations, caveats and comparisons to similar datasets & \textit{Intended Performance, Unintended Application, Unexpected Performance, Caveats, Extended Use Cases, Safety of Use, Downstream Outcomes, Use \& Use Case Evaluation} \\
  \hline
 \textbf{N=1 (examples)} & Examples in the dataset, including typical, outlier, raw and transformed examples; concrete examples or links to additional artifacts of relevance & \textit{Examples or links to typical examples and outliers ; Examples that yield errors; Examples that demonstrate handling of null or zero feature values; code blocks \& scripts, extended documentations, web demos}\\
 \hline
\end{tabular}
\end{table}

\subsection{Evaluation of Data Cards}

To understand how Data Cards were created and maintained, we worked with over 18 producers to understand workflows. 
We found that producers had a tendency to duplicate completed Data Cards (which described similar datasets) as a base template instead of using the provided template. While this made the Data Card easier to complete, it resulted in identical but inaccurate responses.
Errors in the original Data Card would propagate in forked version.
Producers would delete blocks and sections that were irrelevant to their dataset, and in specific cases, producers would semantically modify questions to suit their datasets. Though justifiable in the context of a single Data Card, these practices led to the subsequent fragmentation of forked Data Cards. Deleted but relevant questions were irrecoverable, and reconciling updates to the original template was labor-intensive. 
Producers resorted to answering "N/A"  when they were unsure of the answer, or when uncertainty was high.
These real-world constraints motivated us to identify mechanisms for assuring the quality of Data Cards, and introduce low-barrier processes across the dataset lifecycle that can be easily adopted by organizations. 

Initially, we assigned two reviewers to each new Data Card created. Reviewers represent job functions typical to agents. Selected reviewers were always unfamiliar with the dataset, but typically fluent in manipulating data or the domain of the dataset.
Despite their expertise, feedback provided on these Data Cards were observational and speculative in nature (\textit{"The first two listed applications are commonly used and should be understood by both practitioners and laypeople, but I'm not sure about [application]}); and often not tactical for producers to incorporate. 
To make feedback actionable and holistic, we worked with a mix of subject matter experts, data reviewers, functional and tactical roles at our company to identify over 98 concepts used to assess datasets and their documentation. 
We removed 13 usability and 8 user-experience related concepts, which are reflected in our objectives. 
We then consolidated the remaining concepts into 20 clusters using affinity mapping. 
Clusters were then classified into five umbrella topics that represent contextual decision-making signals used by our experts to evaluate the rigor with which a Data Card describes a dataset, and it's corresponding efficacy for the reader. 

\subsubsection{Dimensions}
Dimensions are directional, pedagogic vectors that describe the Data Card’s usefulness to the agents. They represent the different types of judgments readers might make, and yield qualitative insights into the consistency, comprehensiveness, utility, and readability of Data Card templates and completed Data Cards alike. Here, we briefly summarize these dimensions:

\begin{itemize}
\small
\item \textbf{Accountability}: Demonstrates adequate ownership, reflection, reasoning, and systematic decision making by dataset producers.
\item \textbf{Utility or Use:} Provides details that satisfy the needs of the readers’ responsible decision-making process to establish the suitability of datasets for their tasks and goals.
\item \textbf{Quality: }Summarizes the rigor, integrity and completeness of the dataset, communicated in a manner that is accessible and understandable to many readers.  
\item \textbf{Impact or Consequences of Use:} Sets expectations for positive and negative outcomes as well as subsequent consequences when using or managing the dataset in suitable contexts.
\item \textbf{Risk and Recommendations:} Makes readers aware of known potential risks and limitations, stemming from provenance, representation, use, or context of use. Provides enough information and alternatives to help readers make responsible trade-offs. 
\end{itemize}

To test these dimensions, reviewers with varying levels of domain and data fluency were asked to use these dimensions when evaluating Data Cards \textit{and} any associated Model Cards.
Reviewers were provided a rubric in which they were to independently rate the completed Data Card on each dimension, selecting from \textit{Poor}, \textit{Borderline}, \textit{Average}, \textit{Good}, and \textit{Outstanding}. In addition, reviewers were asked to provide evidence in support of their ratings, and steps that producers could take to improve that specific rating.
Reviewers found it easier to offer structured and actionable feedback using these dimensions (\textit{"Utility or Use: Average. Evidence: Data Card provides all necessary steps for users who may wish to access the dataset, but it's hard for me to determine what use cases are suitable for this dataset. I know the dataset was collected for the purpose of evaluating the performance of the [specific model], but what does the [specific model] do? Next Steps: Provide additional examples of suitable use cases, provide additional detail on what the [specific model] does under intended use case."}). While these dimensions are primarily used to asses if Data Cards help readers arrive at acceptable conclusions about datasets, feedback from expert reviewers revealed specific opportunities to enhance the datasets themselves. 

\section{Case Studies}
\subsection{A Computer Vision Dataset for Fairness Research} \label{cs1}
A research team created an ML training dataset for computer vision (CV) fairness techniques that described sensitive attributes about people, such as perceived gender and perceived age-range. Using Open Images \cite{kuznetsova2020open}, the dataset included 100,000 bounding boxes over 30,000 images. Each bounding box was manually annotated with perceived gender presentation and perceived age range presentation attributes. Given the risks associated with sensitive labels describing personal attributes weighed against the societal benefit of these labels for fairness analysis and bias mitigation, the team wanted an efficient way to provide an overview of the characteristics, limitations, and communicate acceptable uses of the dataset for internal ethics reviewers and external audiences.

Three parties were involved in the creation of this Data Card \cite{chumann_ricco_prabhu_ferrari_pantofaru_2021}, which started after the dataset was completely prepared. First, the dataset authors who had deep tacit knowledge of the processes and decisions across the dataset's lifecycle. They also had explicit knowledge from extensive analysis performed for the dataset release. However, this was distributed across several documents, and the Data Card was an exercise in organizing knowledge into a “readable format” that could be consistently repeated for multiple datasets. This process occurred asynchronously over a few days. 

The next group involved were internal reviewers of the dataset and an accompanying paper, conducting an analysis of how the dataset aligns with responsible AI research and development practices. The analysis focused on subgroups in the labels, the trade-offs associated with each subgroup, and clarifying acceptable and unacceptable use cases of the dataset as a whole, in alignment with an established set of AI Principles. The reviewers recommended that the team create a Data Card. Creating the Data Card as a result of the review process revealed differences in perception across experts. For example, in the Data Card, producers noted that nearly 40\% of perceived age-range labels were 'unknown'. Reviewers were unable to ascertain if this was acceptable, and subsequent conversations raised further questions about the criteria used to label a bounding box with 'unknown' perceived age-range. It was found that 'high' levels of unknowns were relatively typical to datasets in this problem space, and was attributed to the size of 30\% of the bounding boxes being less than 1\% of the image. As a result, producers added a custom section about bounding boxes to the Data Card, and created additional supporting visualizations. Further, producers uncovered and iterated on additional Data Card fields for future CV datasets.

The last group involved in the creation of the Data Card were the authors of this paper, who provided human-centered design perspectives on the Data Card. Feedback was primarily geared towards uncovering agent information needs for acceptable conclusions about the accountability, risk \& recommendations, uses, consequences, and quality of the dataset. A post-launch retrospective revealed that though the producers did not have access to dataset consumers, downstream agents reported finding the Data Card useful, and sought out templates for their own use.

\subsection{A Geographically Diverse Dataset for Language Translation} \label{cs2}
A team of software engineers and a product manager noticed that certain models were “picking up” names to define a person’s gender. Upon investigation, they found that previous training datasets did not have sufficient names that were uncommon in English or belonged to a non-American geography. They also found that model creators were making assumptions about these datasets. In response, the team decided to create a geographically diverse dataset from a limited set of publicly curated data from Wikipedia.

However, it became clear that a truly diverse dataset would need to consider race, age, gender, background and profession as well. While countries were acceptable proxies for geographic representation, gender would need to be inferred from the entity descriptions. Without an awareness of the goals of the dataset or the definitions of gender in the data design, the team was concerned that model creators could make assumptions leading to inappropriate dataset use. To communicate these two aspects, the team created a Data Card for readers with and without technical expertise.

Experts responsible for the design, data extraction, cleaning and curation of the dataset worked with a human-centered designer in an iterative process to produce the Data Card \cite{austermann_linch_stella_webster_2021} after the dataset was created. While the documentation process itself took approximately 20 hours, the Data Card prompted the team to reflect on how data was selected, reviewed and created. They specifically considered what they did and did not know about the dataset, their assumptions, the advantages and limitations of the dataset. In doing so, the team was forced to rethink design decisions which increased the overall timeline, but resulted in a more principled and intentional dataset of geographically diverse biographies. 

The team utilized the Data Card to engage in overall clearer discussions with stakeholders. In particular, experts stakeholders pointed out that gender is difficult to ascertain in the dataset. These conversations helped the team agree on a definition of perceived gender that relied on gender-indicative terms within the text of the data, using the labels “masculine”, “feminine”, and “neutral” for biographies describing collections of individuals. The team found that some discussions around the Data Card were actually about the dataset, and noted the usefulness of this feedback if received during the design stage. The final Data Card describes the data selection criteria, sampling criteria, sources of fields, and emphasizes the distribution of countries by continental regions. In addition, the team was able to clearly justify reasons for not including non-binary individuals, excluding collected data, and the limitations of this dataset. 

\section{Discussion}

\subsubsection{Experiences and outcomes from Case Studies}
While both teams appreciated the transparency added to their respective datasets, creating Data Cards as a final step significantly increased the perception of work required. Rather than a post-implementation task, creating Data Cards as the dataset is created offers several benefits. First, it enables the inclusion of multiple perspectives (engineering, research, user experience, legal, and ethical) to enhance the readability and relevance of documentation, and the dataset quality over time. Then, it forces the aggregation of disparate documentation across the dataset lifecycle into a single, ground truth document accessible to stakeholders. Lastly, it facilitates early feedback on responsible AI practices from experts and non-experts that can affect data design and analyses. Of note, teams that developed multiple Data Cards over a period started developing a nuanced vocabulary to express uncertainty that accurately reflected the status of the information.

\subsubsection{Data Cards as Boundary Objects}
 Data Cards are designed to embody a high degree of interpretive flexibility \cite{leigh2010not}. A single Data Card can support tasks such as conducting reviews and audits, determining use in AI systems or research, comparison of multiple datasets, reproduction of research, or tracking of dataset adoption by various groups. For example, data practitioners seeking to evaluate the quality of a dataset for benchmarking or analysis; AI practitioners determining use case suitability of a dataset for deployment in new or existing models; product managers assessing the downstream effects to make data-related decisions about model or product optimizations for the desired user experience; policy stakeholders evaluating the representativeness of a dataset in relation to end users, and the role of various agencies involved in the creating the dataset creation. Importantly, while Data Cards are able to hold a common identity across these groups, they allow stakeholders to analytically make decisions using dimensions, constructs and vocabulary that are meaningful to their own communities of practice. Data Cards are able to facilitate collaborative work across stakeholders, while supporting individual decision making without consensus.

Our design of Data Cards enables the embedding of relevant sections into transparency artifacts that describe ML models and AI systems. Conversely, sections in the Data Card are designed to capture documentation surrounding the use of datasets in ML models. This establishes a network of artifacts that stakeholders can examine when conducting fairness and accountability interrogations, and achieve overall better results for meta-problems across the domain such as knowledge transfer, dataset reusability, organizational governance, and oversight mechanisms. Data Cards, therefore, effectively act as boundary objects \cite{star1989institutional} and where relevant, boundary infrastructures. 

\subsubsection{Path to Adoption}
Following our initial Data Card release \cite{oix}, public and private organizations have since sought to adopt similar constructs (\cite{gembenchmark}, \cite{huggingfacecards}, \cite{jaic}). Within our organization, we observed an increase in \textit{non-mandated} Data Cards created by individuals who organically came across completed Data Cards. 
While these speak to the utility of Data Cards as a documentation artifact, its quality and comprehensiveness depend on the rigor of the producers, the nuance in expressing uncertainty, and their knowledge of the dataset. Organizational factors include the presence of minimum or mandatory content requirements, process incentives, training materials, and infrastructure for creating and sharing Data Cards. While we propose a relatively comprehensive template for documenting datasets in Data Cards, industry-wide adoption could be spurred by agreed-upon interoperability and content standards that serve as a means for producers and agents to develop more equitable mental models of datasets.

\subsubsection{Infrastructure and Automation}
Critical to an organization's success is its ability to tailor Data Cards to their datasets, models, and technological stack. Knowledge management infrastructures must be connected to data and model pipelines so new knowledge can be seamlessly incorporated into the Data Card, keeping it up to date. We find that Blocks allows for easy implementation on interactive platforms (digital forms, repositories, dataset catalogs) and adaptation for non-interactive surfaces (PDFs, documents, physical papers, markdown files). While both these case studies produced static PDFs, sections and fields can be easily implemented in a browser-based user interface, configured for views tailored to different stakeholders.

Centralized repositories that can perform search-and-filter operations over hundreds of Data Cards have long-tail benefits for agents in identifying the most suitable datasets for their tasks; measurably distributing the accountability of how datasets are used. We observed a marked preference for infrastructures that enables stakeholder collaboration and co-creation of Data Cards, linking and storage of extraneous artifacts, and the partial automation of visualizations, tables and analyses results. Interestingly, we observed that readers had strong opinions about \textit{not} automating certain fields in the Data Card, especially when responses contain assumptions or rationales that help interpret results. Fields should be automated to guarantee accuracy and antifragility at all times, preventing the misrepresentation and the subsequent legitimizing of poor quality datasets. Implicit knowledge is articulated by providing contextual, human-written explanations of methods, assumptions, decisions and baselines.

\section{Conclusion}
We present Data Cards, a framework for transparent and purposeful documentation of datasets. We describe our underlying approach, objectives and the human-centered design of Data Cards. We introduce frameworks for structuring, adapting, and evaluating Data Cards. 
We then detail qualitative and anecdotal evidence for the efficacy of Data Cards towards creating responsible AI systems through two case studies.
A limitation of our approach was the use of Google Docs for Data Card templates. This allowed stakeholders to collaborate and preserved a forensic history of the development of the Data Card, producers were limited to providing answers using text, tables and images. 
Iterating on individual fields caused template fragmentation and the loss of the original intent, as observed in our first case study. Additionally, this format prevented us from streamlining the collection experience or introducing automations, a much requested feature from producers. 
Future work requires a more principled approach for extending and adapting Data Card templates without compromising comparability. 
Insights from studies call for participatory approaches that engage diverse, non-traditional stakeholders early into the dataset \textit{and} Data Card development process. 
We believe that adopting a co-creative approach that spans the entire dataset life cycle will result in a deliberate approach to automation in documentation.
Lastly, defining quantitative measures to assess the true value of Data Cards will require adoption at both breadth and depth in the industry. 
Further investigation is needed into the perceived and actual importance of the content of Data Cards to tasks for different stakeholder groups. 
Together, Data Cards templates and frameworks encourage customized implementations that foster a culture for deep, detailed documentation. These help produce Data Cards that are consistent not only in and of themselves, but also co-create practical industry standards. Data Cards are capable of thoughtfully explaining the implications of datasets while highlighting unknowns appropriately. They reveals insights about inherent aspects of dataset that cannot be intrinsically determined by interacting with the dataset. 
Data Cards can be powerful vehicles that emphasize the ethical considerations of a dataset in ways that can be practically acted upon and support production and research decisions, supporting transparent and well-informed development of large AI models capable of supporting multiple, user-facing tasks.



\bibliographystyle{ACM-Reference-Format}
\bibliography{references.bib}


\begin{thebibliography}{26}


\ifx \showCODEN    \undefined \def \showCODEN     #1{\unskip}     \fi
\ifx \showDOI      \undefined \def \showDOI       #1{#1}\fi
\ifx \showISBNx    \undefined \def \showISBNx     #1{\unskip}     \fi
\ifx \showISBNxiii \undefined \def \showISBNxiii  #1{\unskip}     \fi
\ifx \showISSN     \undefined \def \showISSN      #1{\unskip}     \fi
\ifx \showLCCN     \undefined \def \showLCCN      #1{\unskip}     \fi
\ifx \shownote     \undefined \def \shownote      #1{#1}          \fi
\ifx \showarticletitle \undefined \def \showarticletitle #1{#1}   \fi
\ifx \showURL      \undefined \def \showURL       {\relax}        \fi
\providecommand\bibfield[2]{#2}
\providecommand\bibinfo[2]{#2}
\providecommand\natexlab[1]{#1}
\providecommand\showeprint[2][]{arXiv:#2}

\bibitem[\protect\citeauthoryear{??}{ain}{2017}]%
        {ainow}
 \bibinfo{year}{2017}\natexlab{}.
\newblock \bibinfo{booktitle}{\emph{AI Now Institute}}.
\newblock
\urldef\tempurl%
\url{https://ainowinstitute.org/}
\showURL{%
\tempurl}


\bibitem[\protect\citeauthoryear{??}{fac}{2021}]%
        {facct}
 \bibinfo{year}{2021}\natexlab{}.
\newblock \bibinfo{booktitle}{\emph{ACM Conference on Fairness, Accountability,
  and Transparency (ACM FAccT)}}.
\newblock
\urldef\tempurl%
\url{https://facctconference.org/}
\showURL{%
\tempurl}


\bibitem[\protect\citeauthoryear{Affairs}{Affairs}{2021}]%
        {jaic}
\bibfield{author}{\bibinfo{person}{Joint Artificial Intelligence Center~Public
  Affairs}.} \bibinfo{year}{2021}\natexlab{}.
\newblock \bibinfo{title}{Enabling AI with Data Cards}.
\newblock
\newblock
\urldef\tempurl%
\url{https://www.ai.mil/blog_09_03_21_ai_enabling_ai_with_data_cards.html}
\showURL{%
\tempurl}


\bibitem[\protect\citeauthoryear{Antunes, Balby, Figueiredo, Lourenco, Meira,
  and Santos}{Antunes et~al\mbox{.}}{2018}]%
        {antunes2018fairness}
\bibfield{author}{\bibinfo{person}{Nuno Antunes}, \bibinfo{person}{Leandro
  Balby}, \bibinfo{person}{Flavio Figueiredo}, \bibinfo{person}{Nuno Lourenco},
  \bibinfo{person}{Wagner Meira}, {and} \bibinfo{person}{Walter Santos}.}
  \bibinfo{year}{2018}\natexlab{}.
\newblock \showarticletitle{Fairness and transparency of machine learning for
  trustworthy cloud services}. In \bibinfo{booktitle}{\emph{2018 48th Annual
  IEEE/IFIP International Conference on Dependable Systems and Networks
  Workshops (DSN-W)}}. IEEE, \bibinfo{pages}{188--193}.
\newblock


\bibitem[\protect\citeauthoryear{Anurag~Batra}{Anurag~Batra}{2020}]%
        {oix}
\bibfield{author}{\bibinfo{person}{Parker~Barnes Anurag~Batra}.}
  \bibinfo{year}{2020}\natexlab{}.
\newblock \bibinfo{title}{Open Images Extended - Crowdsourced Data Card}.
\newblock
\newblock
\urldef\tempurl%
\url{https://research.google/static/documents/datasets/open-images-extended-crowdsourced.pdf}
\showURL{%
\tempurl}


\bibitem[\protect\citeauthoryear{Austermann, Linch, Stella, and
  Webster}{Austermann et~al\mbox{.}}{2021}]%
        {austermann_linch_stella_webster_2021}
\bibfield{author}{\bibinfo{person}{Anja Austermann}, \bibinfo{person}{Michelle
  Linch}, \bibinfo{person}{Romina Stella}, {and} \bibinfo{person}{Kellie
  Webster}.} \bibinfo{year}{2021}\natexlab{}.
\newblock
\newblock
\urldef\tempurl%
\url{https://storage.googleapis.com/gresearch/translate-gender-challenge-sets/Data\%20Card.pdf}
\showURL{%
\tempurl}


\bibitem[\protect\citeauthoryear{Barclay, Taylor, Preece, Taylor, Verma, and
  de~Mel}{Barclay et~al\mbox{.}}{2020}]%
        {barclay2020framework}
\bibfield{author}{\bibinfo{person}{Iain Barclay}, \bibinfo{person}{Harrison
  Taylor}, \bibinfo{person}{Alun Preece}, \bibinfo{person}{Ian Taylor},
  \bibinfo{person}{Dinesh Verma}, {and} \bibinfo{person}{Geeth de Mel}.}
  \bibinfo{year}{2020}\natexlab{}.
\newblock \showarticletitle{A framework for fostering transparency in shared
  artificial intelligence models by increasing visibility of contributions}.
\newblock \bibinfo{journal}{\emph{Concurrency and Computation: Practice and
  Experience}} (\bibinfo{year}{2020}), \bibinfo{pages}{e6129}.
\newblock


\bibitem[\protect\citeauthoryear{Bender and Friedman}{Bender and
  Friedman}{2018}]%
        {bender2018data}
\bibfield{author}{\bibinfo{person}{Emily~M Bender} {and} \bibinfo{person}{Batya
  Friedman}.} \bibinfo{year}{2018}\natexlab{}.
\newblock \showarticletitle{Data statements for natural language processing:
  Toward mitigating system bias and enabling better science}.
\newblock \bibinfo{journal}{\emph{Transactions of the Association for
  Computational Linguistics}}  \bibinfo{volume}{6} (\bibinfo{year}{2018}),
  \bibinfo{pages}{587--604}.
\newblock


\bibitem[\protect\citeauthoryear{Bhatt, Antor{\'a}n, Zhang, Liao, Sattigeri,
  Fogliato, Melan{\c{c}}on, Krishnan, Stanley, Tickoo, et~al\mbox{.}}{Bhatt
  et~al\mbox{.}}{2021}]%
        {bhatt2021uncertainty}
\bibfield{author}{\bibinfo{person}{Umang Bhatt}, \bibinfo{person}{Javier
  Antor{\'a}n}, \bibinfo{person}{Yunfeng Zhang}, \bibinfo{person}{Q~Vera Liao},
  \bibinfo{person}{Prasanna Sattigeri}, \bibinfo{person}{Riccardo Fogliato},
  \bibinfo{person}{Gabrielle Melan{\c{c}}on}, \bibinfo{person}{Ranganath
  Krishnan}, \bibinfo{person}{Jason Stanley}, \bibinfo{person}{Omesh Tickoo},
  {et~al\mbox{.}}} \bibinfo{year}{2021}\natexlab{}.
\newblock \showarticletitle{Uncertainty as a form of transparency: Measuring,
  communicating, and using uncertainty}. In
  \bibinfo{booktitle}{\emph{Proceedings of the 2021 AAAI/ACM Conference on AI,
  Ethics, and Society}}. \bibinfo{pages}{401--413}.
\newblock


\bibitem[\protect\citeauthoryear{Chander, Srinivasan, Chelian, Wang, and
  Uchino}{Chander et~al\mbox{.}}{2018}]%
        {chander2018working}
\bibfield{author}{\bibinfo{person}{Ajay Chander}, \bibinfo{person}{Ramya
  Srinivasan}, \bibinfo{person}{Suhas Chelian}, \bibinfo{person}{Jun Wang},
  {and} \bibinfo{person}{Kanji Uchino}.} \bibinfo{year}{2018}\natexlab{}.
\newblock \showarticletitle{Working with beliefs: AI transparency in the
  enterprise}. In \bibinfo{booktitle}{\emph{IUI Workshops}}.
\newblock


\bibitem[\protect\citeauthoryear{chumann, Ricco, Prabhu, Ferrari, and
  Pantofaru}{chumann et~al\mbox{.}}{2021}]%
        {chumann_ricco_prabhu_ferrari_pantofaru_2021}
\bibfield{author}{\bibinfo{person}{Candice chumann}, \bibinfo{person}{Susanna
  Ricco}, \bibinfo{person}{Utsav Prabhu}, \bibinfo{person}{Vittorio Ferrari},
  {and} \bibinfo{person}{Caroline Pantofaru}.} \bibinfo{year}{2021}\natexlab{}.
\newblock
\newblock
\urldef\tempurl%
\url{https://storage.googleapis.com/openimages/open_images_extended_miap/Open\%20Images\%20Extended\%20-\%20MIAP\%20-\%20Data\%20Card.pdf}
\showURL{%
\tempurl}


\bibitem[\protect\citeauthoryear{Ehsan, Liao, Muller, Riedl, and Weisz}{Ehsan
  et~al\mbox{.}}{2021}]%
        {ehsan2021expanding}
\bibfield{author}{\bibinfo{person}{Upol Ehsan}, \bibinfo{person}{Q~Vera Liao},
  \bibinfo{person}{Michael Muller}, \bibinfo{person}{Mark~O Riedl}, {and}
  \bibinfo{person}{Justin~D Weisz}.} \bibinfo{year}{2021}\natexlab{}.
\newblock \showarticletitle{Expanding explainability: Towards social
  transparency in ai systems}. In \bibinfo{booktitle}{\emph{Proceedings of the
  2021 CHI Conference on Human Factors in Computing Systems}}.
  \bibinfo{pages}{1--19}.
\newblock


\bibitem[\protect\citeauthoryear{Felzmann, Fosch-Villaronga, Lutz, and
  Tam{\`o}-Larrieux}{Felzmann et~al\mbox{.}}{2020}]%
        {felzmann2020towards}
\bibfield{author}{\bibinfo{person}{Heike Felzmann}, \bibinfo{person}{Eduard
  Fosch-Villaronga}, \bibinfo{person}{Christoph Lutz}, {and}
  \bibinfo{person}{Aurelia Tam{\`o}-Larrieux}.}
  \bibinfo{year}{2020}\natexlab{}.
\newblock \showarticletitle{Towards transparency by design for artificial
  intelligence}.
\newblock \bibinfo{journal}{\emph{Science and Engineering Ethics}}
  \bibinfo{volume}{26}, \bibinfo{number}{6} (\bibinfo{year}{2020}),
  \bibinfo{pages}{3333--3361}.
\newblock


\bibitem[\protect\citeauthoryear{Gebru, Morgenstern, Vecchione, Vaughan,
  Wallach, Daum{\'e}~III, and Crawford}{Gebru et~al\mbox{.}}{2018}]%
        {gebru2018datasheets}
\bibfield{author}{\bibinfo{person}{Timnit Gebru}, \bibinfo{person}{Jamie
  Morgenstern}, \bibinfo{person}{Briana Vecchione},
  \bibinfo{person}{Jennifer~Wortman Vaughan}, \bibinfo{person}{Hanna Wallach},
  \bibinfo{person}{Hal Daum{\'e}~III}, {and} \bibinfo{person}{Kate Crawford}.}
  \bibinfo{year}{2018}\natexlab{}.
\newblock \showarticletitle{Datasheets for datasets}.
\newblock \bibinfo{journal}{\emph{arXiv preprint arXiv:1803.09010}}
  (\bibinfo{year}{2018}).
\newblock


\bibitem[\protect\citeauthoryear{GEM}{GEM}{2022}]%
        {gembenchmark}
\bibfield{author}{\bibinfo{person}{GEM}.} \bibinfo{year}{2022}\natexlab{}.
\newblock \bibinfo{title}{Natural Language Generation, its Evaluation and
  Metrics Data Cards}.
\newblock
\newblock
\urldef\tempurl%
\url{https://gem-benchmark.com/data_cards}
\showURL{%
\tempurl}


\bibitem[\protect\citeauthoryear{HuggingFace}{HuggingFace}{2021}]%
        {huggingfacecards}
\bibfield{author}{\bibinfo{person}{HuggingFace}.}
  \bibinfo{year}{2021}\natexlab{}.
\newblock \bibinfo{title}{HuggingFace - Create a Dataset Card}.
\newblock
\newblock
\urldef\tempurl%
\url{https://huggingface.co/docs/datasets/v1.12.0/dataset_card.html}
\showURL{%
\tempurl}


\bibitem[\protect\citeauthoryear{Hutchinson, Smart, Hanna, Denton, Greer,
  Kjartansson, Barnes, and Mitchell}{Hutchinson et~al\mbox{.}}{2021}]%
        {hutchinson2021towards}
\bibfield{author}{\bibinfo{person}{Ben Hutchinson}, \bibinfo{person}{Andrew
  Smart}, \bibinfo{person}{Alex Hanna}, \bibinfo{person}{Emily Denton},
  \bibinfo{person}{Christina Greer}, \bibinfo{person}{Oddur Kjartansson},
  \bibinfo{person}{Parker Barnes}, {and} \bibinfo{person}{Margaret Mitchell}.}
  \bibinfo{year}{2021}\natexlab{}.
\newblock \showarticletitle{Towards accountability for machine learning
  datasets: Practices from software engineering and infrastructure}. In
  \bibinfo{booktitle}{\emph{Proceedings of the 2021 ACM Conference on Fairness,
  Accountability, and Transparency}}. \bibinfo{pages}{560--575}.
\newblock


\bibitem[\protect\citeauthoryear{Initiative}{Initiative}{2022}]%
        {kyd}
\bibfield{author}{\bibinfo{person}{People + AI~Research Initiative}.}
  \bibinfo{year}{2022}\natexlab{}.
\newblock \bibinfo{title}{Know Your Data}.
\newblock
\newblock
\urldef\tempurl%
\url{https://knowyourdata.withgoogle.com/}
\showURL{%
\tempurl}


\bibitem[\protect\citeauthoryear{Kuznetsova, Rom, Alldrin, Uijlings, Krasin,
  Pont-Tuset, Kamali, Popov, Malloci, Kolesnikov, et~al\mbox{.}}{Kuznetsova
  et~al\mbox{.}}{2020}]%
        {kuznetsova2020open}
\bibfield{author}{\bibinfo{person}{Alina Kuznetsova}, \bibinfo{person}{Hassan
  Rom}, \bibinfo{person}{Neil Alldrin}, \bibinfo{person}{Jasper Uijlings},
  \bibinfo{person}{Ivan Krasin}, \bibinfo{person}{Jordi Pont-Tuset},
  \bibinfo{person}{Shahab Kamali}, \bibinfo{person}{Stefan Popov},
  \bibinfo{person}{Matteo Malloci}, \bibinfo{person}{Alexander Kolesnikov},
  {et~al\mbox{.}}} \bibinfo{year}{2020}\natexlab{}.
\newblock \showarticletitle{The open images dataset v4}.
\newblock \bibinfo{journal}{\emph{International Journal of Computer Vision}}
  \bibinfo{volume}{128}, \bibinfo{number}{7} (\bibinfo{year}{2020}),
  \bibinfo{pages}{1956--1981}.
\newblock


\bibitem[\protect\citeauthoryear{Leigh~Star}{Leigh~Star}{2010}]%
        {leigh2010not}
\bibfield{author}{\bibinfo{person}{Susan Leigh~Star}.}
  \bibinfo{year}{2010}\natexlab{}.
\newblock \showarticletitle{This is not a boundary object: Reflections on the
  origin of a concept}.
\newblock \bibinfo{journal}{\emph{Science, Technology, \& Human Values}}
  \bibinfo{volume}{35}, \bibinfo{number}{5} (\bibinfo{year}{2010}),
  \bibinfo{pages}{601--617}.
\newblock


\bibitem[\protect\citeauthoryear{McCue}{McCue}{2014}]%
        {mccue2014data}
\bibfield{author}{\bibinfo{person}{Colleen McCue}.}
  \bibinfo{year}{2014}\natexlab{}.
\newblock \bibinfo{booktitle}{\emph{Data mining and predictive analysis:
  Intelligence gathering and crime analysis}}.
\newblock \bibinfo{publisher}{Butterworth-Heinemann}.
\newblock


\bibitem[\protect\citeauthoryear{Mitchell, Wu, Zaldivar, Barnes, Vasserman,
  Hutchinson, Spitzer, Raji, and Gebru}{Mitchell et~al\mbox{.}}{2019}]%
        {mitchell2019model}
\bibfield{author}{\bibinfo{person}{Margaret Mitchell}, \bibinfo{person}{Simone
  Wu}, \bibinfo{person}{Andrew Zaldivar}, \bibinfo{person}{Parker Barnes},
  \bibinfo{person}{Lucy Vasserman}, \bibinfo{person}{Ben Hutchinson},
  \bibinfo{person}{Elena Spitzer}, \bibinfo{person}{Inioluwa~Deborah Raji},
  {and} \bibinfo{person}{Timnit Gebru}.} \bibinfo{year}{2019}\natexlab{}.
\newblock \showarticletitle{Model cards for model reporting}. In
  \bibinfo{booktitle}{\emph{Proceedings of the conference on fairness,
  accountability, and transparency}}. \bibinfo{pages}{220--229}.
\newblock


\bibitem[\protect\citeauthoryear{Pushkarna, Zaldivar, and Nanas}{Pushkarna
  et~al\mbox{.}}{[n.\,d.]}]%
        {craftplaybook}
\bibfield{author}{\bibinfo{person}{Mahima Pushkarna}, \bibinfo{person}{Andrew
  Zaldivar}, {and} \bibinfo{person}{Daniel Nanas}.}
  \bibinfo{year}{[n.\,d.]}\natexlab{}.
\newblock \bibinfo{title}{Data Cards Playbook: Participatory Activities for
  Dataset Documentation}.
\newblock
\newblock
\urldef\tempurl%
\url{https://facctconference.org/2021/acceptedcraftsessions.html#data_cards}
\showURL{%
\tempurl}


\bibitem[\protect\citeauthoryear{Shneiderman}{Shneiderman}{2003}]%
        {shneiderman2003eyes}
\bibfield{author}{\bibinfo{person}{Ben Shneiderman}.}
  \bibinfo{year}{2003}\natexlab{}.
\newblock \showarticletitle{The eyes have it: A task by data type taxonomy for
  information visualizations}.
\newblock In \bibinfo{booktitle}{\emph{The craft of information
  visualization}}. \bibinfo{publisher}{Elsevier}, \bibinfo{pages}{364--371}.
\newblock


\bibitem[\protect\citeauthoryear{Star and Griesemer}{Star and
  Griesemer}{1989}]%
        {star1989institutional}
\bibfield{author}{\bibinfo{person}{Susan~Leigh Star} {and}
  \bibinfo{person}{James~R Griesemer}.} \bibinfo{year}{1989}\natexlab{}.
\newblock \showarticletitle{Institutional ecology,translations' and boundary
  objects: Amateurs and professionals in Berkeley's Museum of Vertebrate
  Zoology, 1907-39}.
\newblock \bibinfo{journal}{\emph{Social studies of science}}
  \bibinfo{volume}{19}, \bibinfo{number}{3} (\bibinfo{year}{1989}),
  \bibinfo{pages}{387--420}.
\newblock


\bibitem[\protect\citeauthoryear{Suresh, Gomez, Nam, and Satyanarayan}{Suresh
  et~al\mbox{.}}{2021}]%
        {suresh2021beyond}
\bibfield{author}{\bibinfo{person}{Harini Suresh}, \bibinfo{person}{Steven~R
  Gomez}, \bibinfo{person}{Kevin~K Nam}, {and} \bibinfo{person}{Arvind
  Satyanarayan}.} \bibinfo{year}{2021}\natexlab{}.
\newblock \showarticletitle{Beyond Expertise and Roles: A Framework to
  Characterize the Stakeholders of Interpretable Machine Learning and their
  Needs}. In \bibinfo{booktitle}{\emph{Proceedings of the 2021 CHI Conference
  on Human Factors in Computing Systems}}. \bibinfo{pages}{1--16}.
\newblock


\end{thebibliography}

\appendix
\newpage
\section{Typology of Stakeholders} \label{appendix-1}

\begin{figure}[!htbp]
    \centering
    \caption{A typology of typical stakeholders in the life cycle of datasets that we created Data Cards for, broken down by type, identifiers and tasks with example roles. We find that including non-technical and indirect stakeholders in a dataset's lifecycle during initial considerations of content and structure builds foresight for successful Data Card adoption.}
   \includegraphics[width=\textwidth]{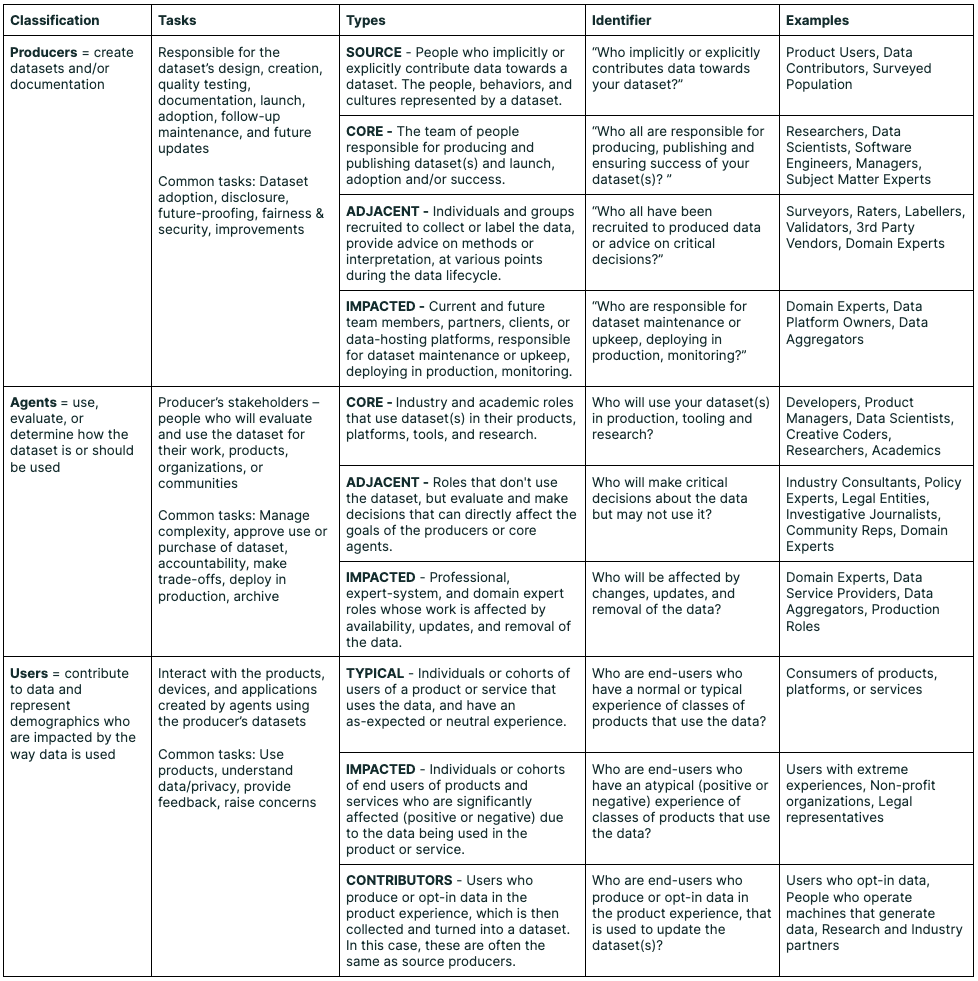}
    \label{typologytable}
\end{figure}

\newpage
\section{OFTEn Framework as a generative tool}

\begin{table}[ht]
\label{often-matrix}
\centering
 \caption{ In this figure, we demonstrate how it can be used to generate questions about data consent across a dataset's life cycle. During the creation of our template (\ref{DC-template}), OFTEn was used to anticipate standardization requirements and enable the forensic investigation of dataset documentation over time. }
\scriptsize
\begin{tabular} {|p{0.04\textwidth}|p{0.16\textwidth}|p{0.16\textwidth}|p{0.16\textwidth}|p{0.16\textwidth}|p{0.16\textwidth}|}
\hline
 & Who & What & When & Where & Why\\
 \hline
 \textbf{O} & Who was responsible for setting the terms of consent? & What were the terms of consent? & When do the terms of consent expire? & Where all are the terms of consent applicable? Are there any exceptions? & Why were these specific terms of consent chosen?\\
 \hline
 \textbf{F}  & How was consent delivered to the surveyed population? & How many data points accompanied consent? & When was the consent collected with respect to data creation or collection? & Where can the consent be accessed? How is it stored? & If at all, why were exceptions made? What happened in cases where consent was not or conditionally provided? Provided but revoked?\\
 \hline
 \textbf{T} & Who tracks consent? & What manipulations of the data are permissible under the given consent? & When can consent be revoked? & X & Why are said transformations in direct conflict with consent?\\
 \hline
 \textbf{E} & Under the terms of the consent, who all can use the dataset? & Under the terms of the consent, what are the permissible uses of the dataset? & When must consent be reacquired from individuals to sustain use of the dataset? & Geographically, where all does the consent permit dataset use?  & Summarize conditions and rationales that justify the use of data without consent. \\
 \hline
 \textbf{N=1} & Provide an example of a consent form & Provide an example of a data point with partial consent & X & X & X\\
 \hline
 \end{tabular}
\end{table}

\newpage
\section{Data Card for Computer Vision Dataset} \label{DC1}

\begin{figure}[!htbp]
    \centering
    \includegraphics[page=1, width=\textwidth, height=0.9\textheight, frame]{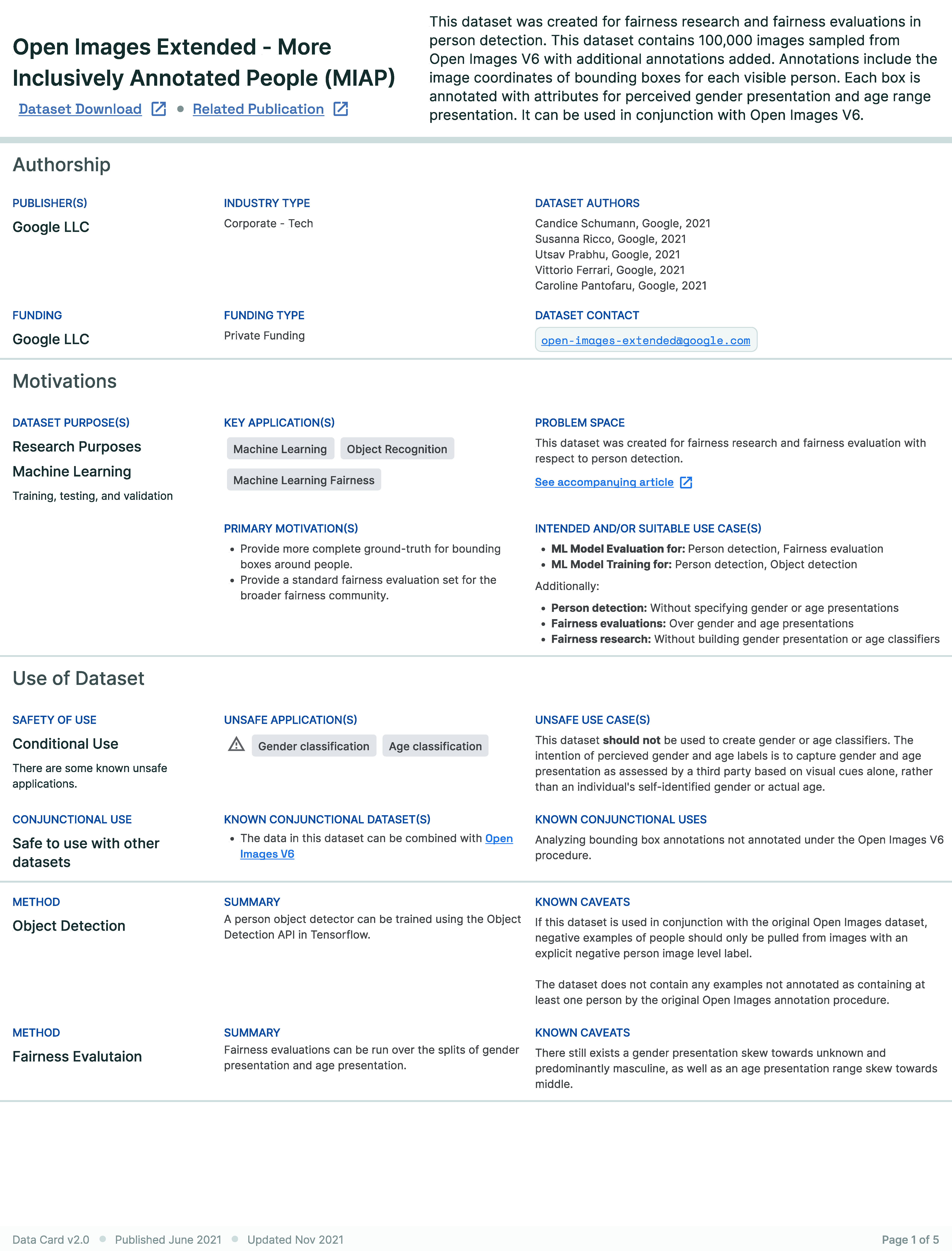}
    \caption{Data Card for Computer Vision Dataset, Page 1 of 5}
    \label{fig:miap-p1}
\end{figure}

\newpage
\begin{figure}[!htbp]
    \ContinuedFloat
    \centering
    \addtocounter{figure}{+1}
    \includegraphics[page=2, width=\textwidth, height=0.9\textheight, frame]{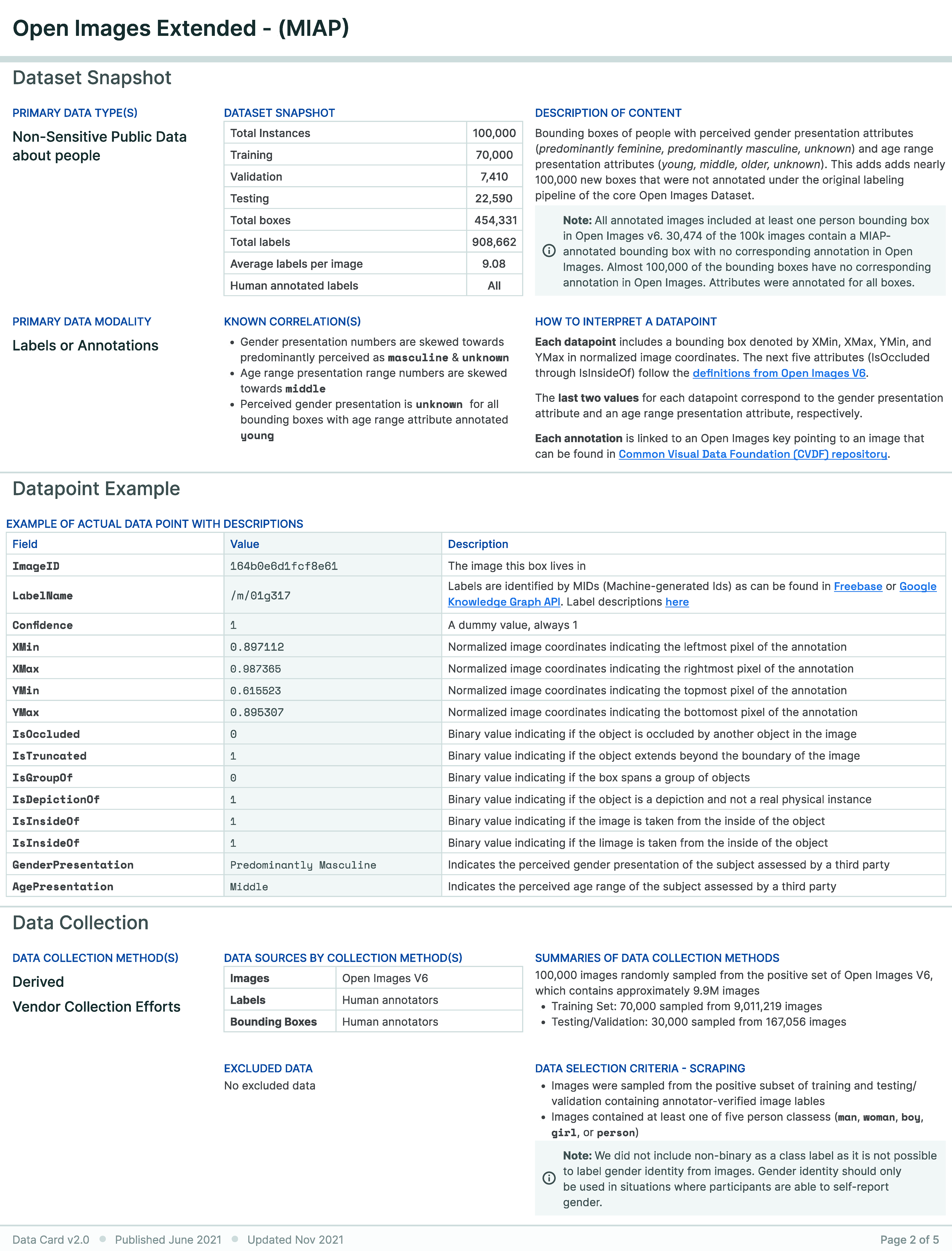}
    \caption{Data Card for Computer Vision Dataset, Page 2 of 5}
    \label{fig:miap-p2}
\end{figure}

\newpage
\begin{figure}[!htbp]
    \ContinuedFloat
    \centering
    \addtocounter{figure}{+1}
    \includegraphics[page=3, width=\textwidth, height=0.9\textheight, frame]{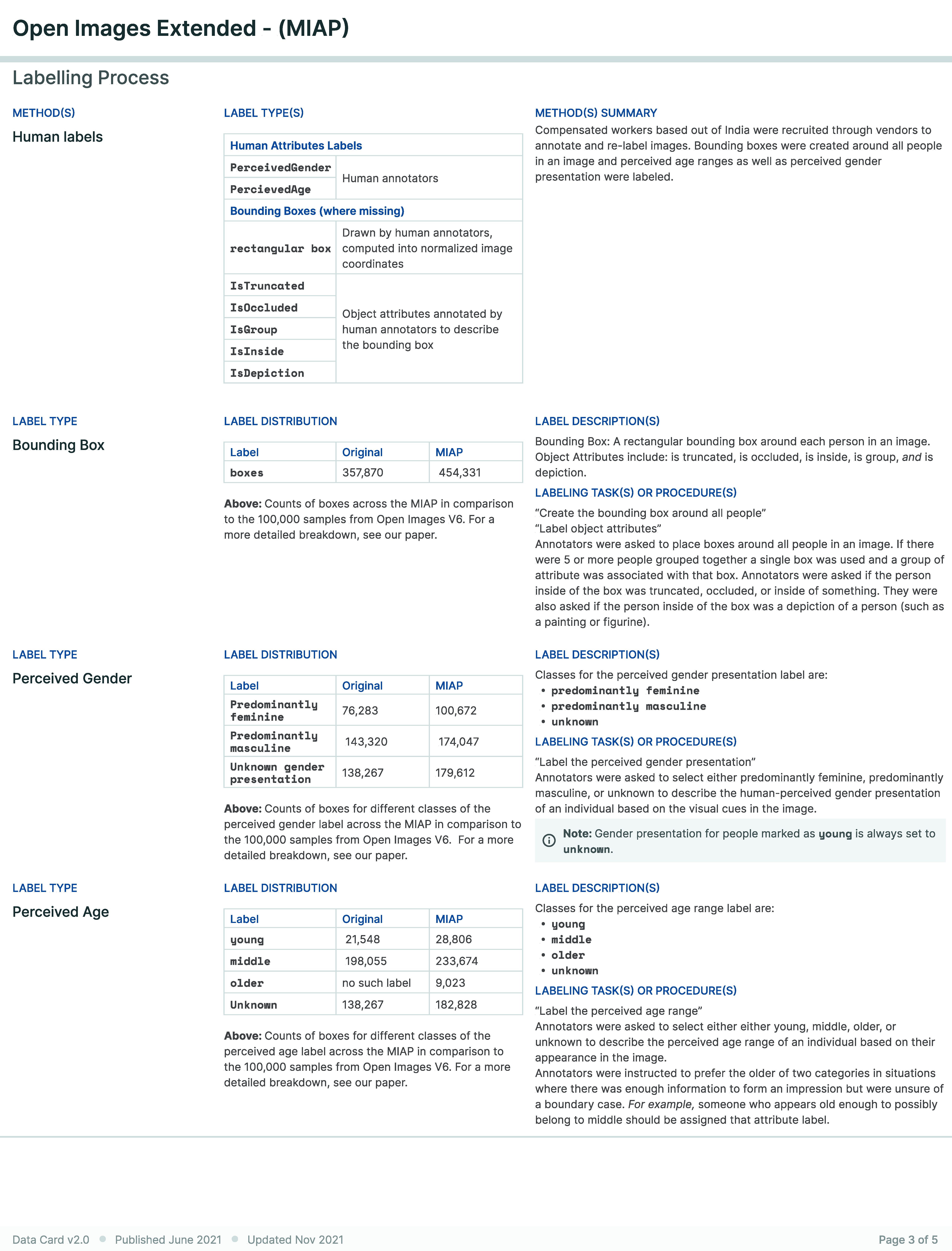}
    \caption{Data Card for Computer Vision Dataset, Page 3 of 5}
    \label{fig:miap-p3}
\end{figure}

\newpage
\begin{figure}[!htbp]
    \ContinuedFloat
    \centering
    \addtocounter{figure}{+1}
  \includegraphics[page=4, width=\textwidth, height=0.9\textheight, frame]{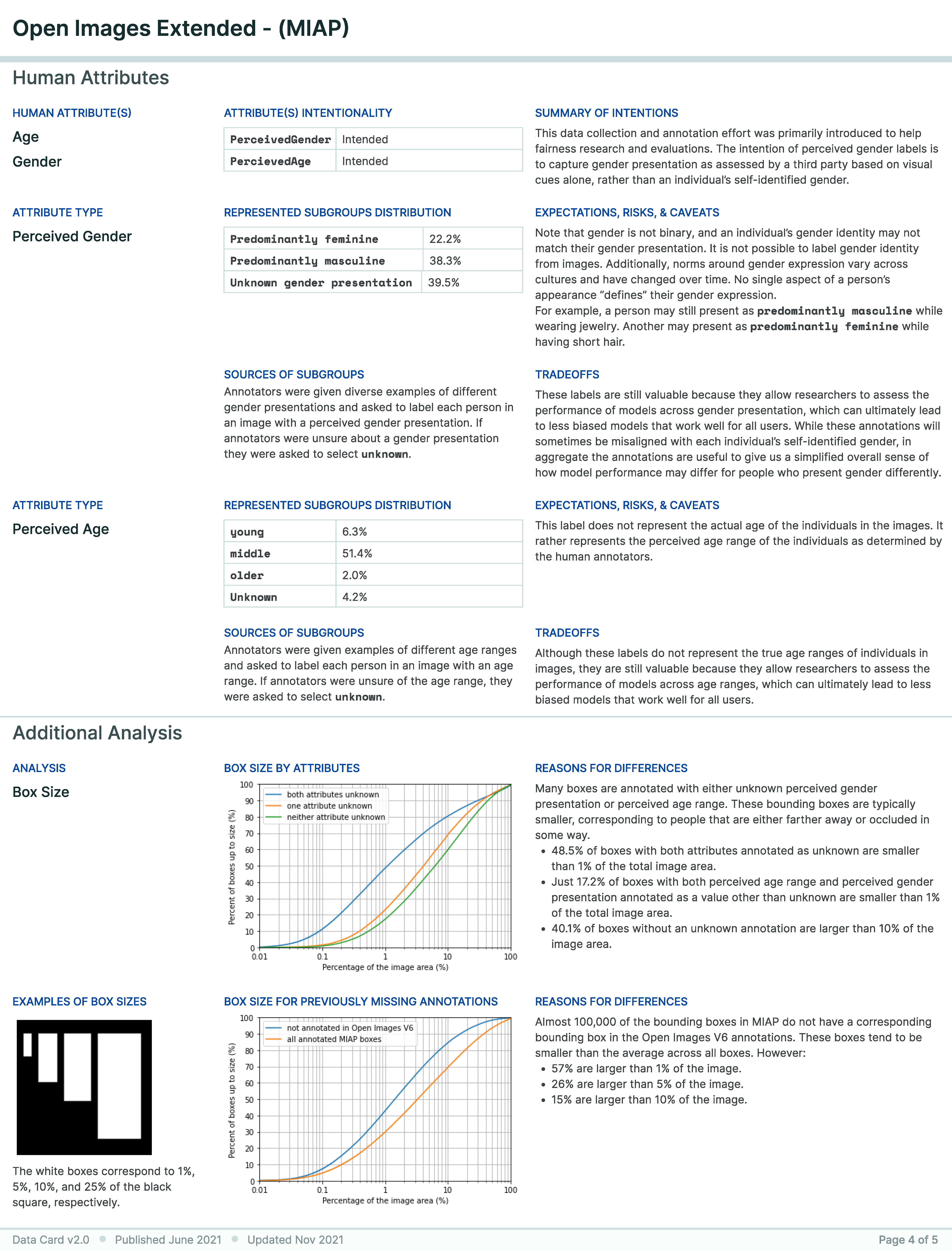}
    \caption{Data Card for Computer Vision Dataset, Page 4 of 5}
    \label{fig:miap-p4}
\end{figure}

\newpage
\begin{figure}[!htbp]
    \ContinuedFloat
    \centering
    \addtocounter{figure}{+1}
    \includegraphics[page=5, width=\textwidth, height=0.9\textheight, frame]{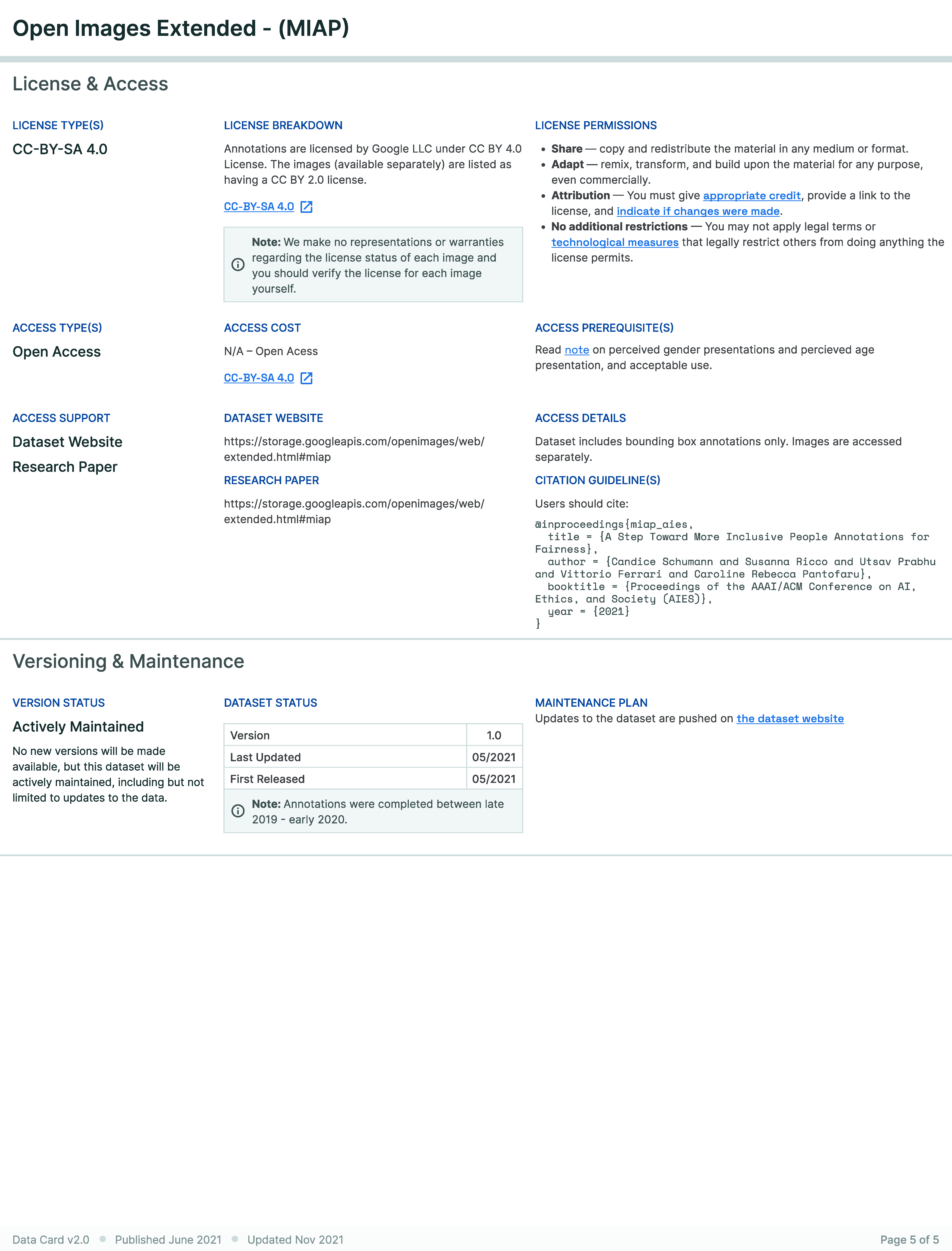}
    \caption{Data Card for Computer Vision Dataset, Page 5 of 5}
    \label{fig:miap-p5}
\end{figure}

\newpage
\section{Data Card for Language Translation Dataset} \label{DC2}
\begin{figure}[!htbp]
    \centering
    \includegraphics[width=\textwidth, height=0.9\textheight, frame]{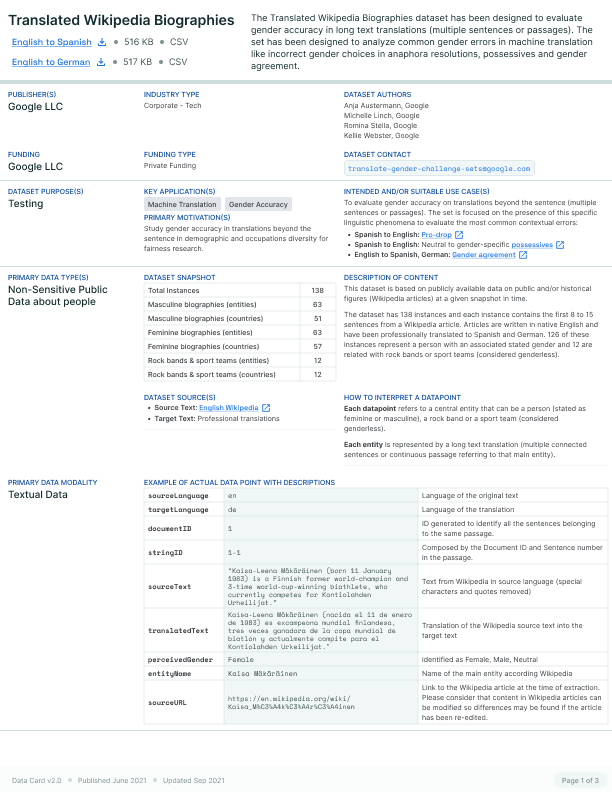}
    \caption{Data Card for Language Translation Dataset, Page 1 of 3}
    \label{fig:twb-p1}
\end{figure}
\begin{figure}[!htbp]
    \centering
    \includegraphics[width=\textwidth, height=0.9\textheight, frame]{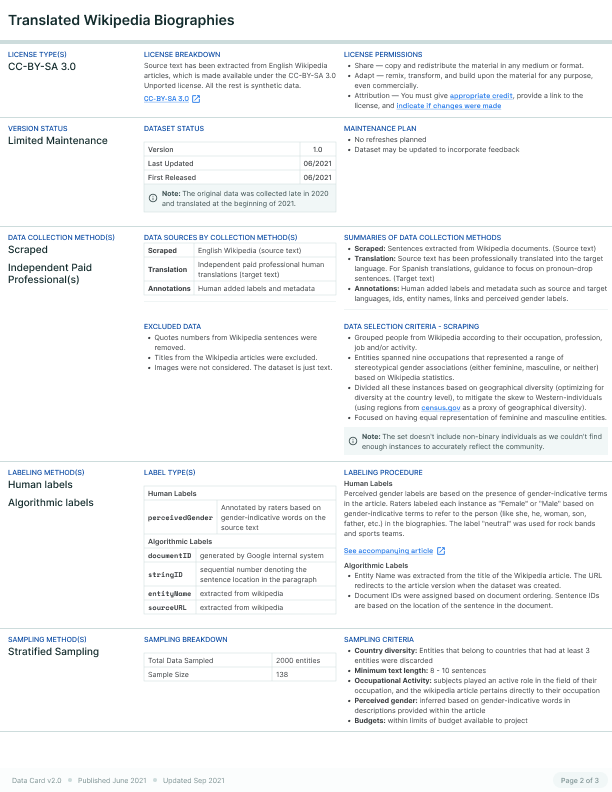}
    \caption{Data Card for Language Translation Dataset, Page 2 of 3}
    \label{fig:twb-p2}
\end{figure}
\begin{figure}[!htbp]
    \centering
    \includegraphics[width=\textwidth, height=0.9\textheight, frame]{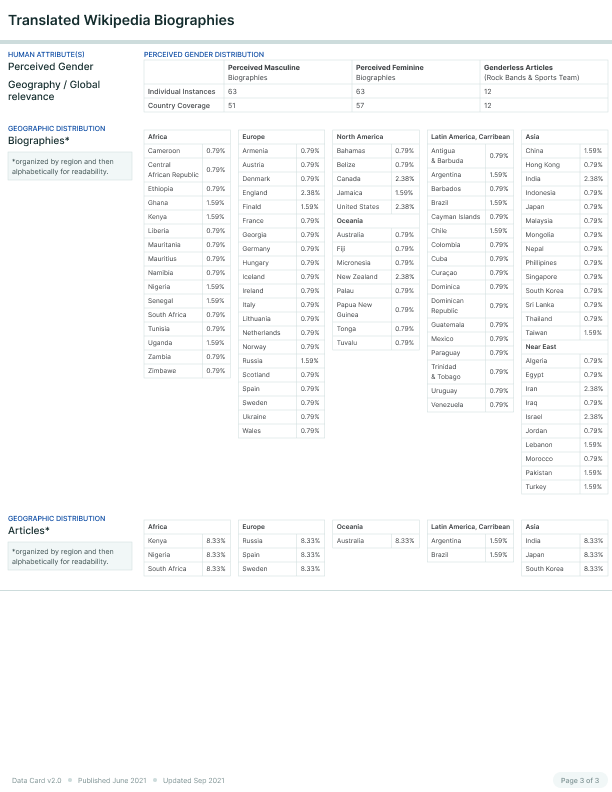}
    \caption{Data Card for Language Translation Dataset, Page 3 of 3}
    \label{fig:twb-p3}
\end{figure}

\section{Data Card Template} \label{DC-template}
\begin{figure}[!htbp]
    \centering
    \includegraphics[width=\textwidth, height=0.85\textheight, frame]{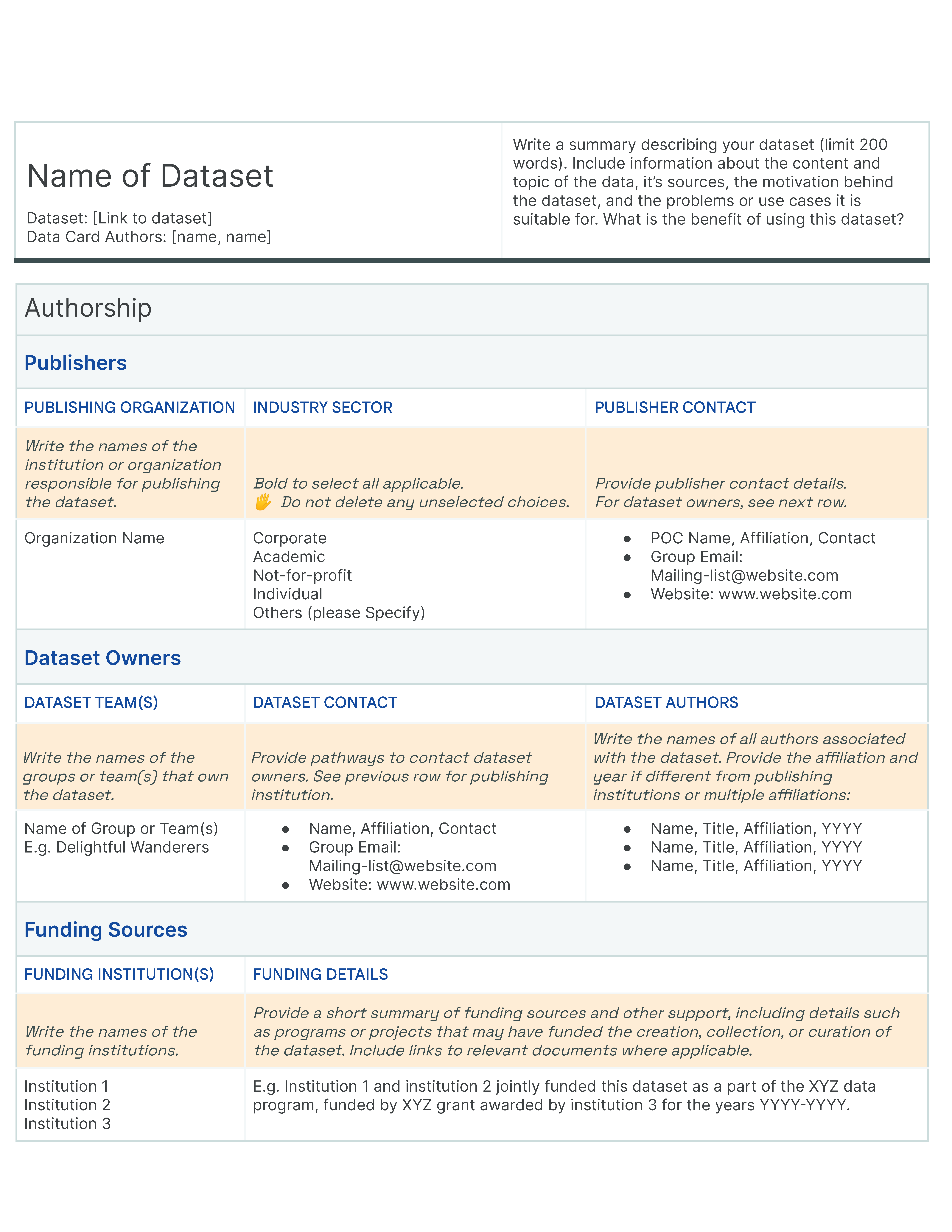}
    \caption{Data Card Template - The \textit{Summary} section introduces the dataset and the authors of the Data Card. The \textit{Authorship} section describes the authors of the dataset. This includes subsections on \textit{Publishers}, which may be different from \textit{Dataset Owners}. The \textit{Funding Sources} subsection describes grants and programs academic, research, and industry organizations that supported the creation of the dataset from.}
    \label{fig:dct-p1}
\end{figure}

\begin{figure}[!htbp]
    \centering
    \includegraphics[width=\textwidth, height=0.9\textheight, frame]{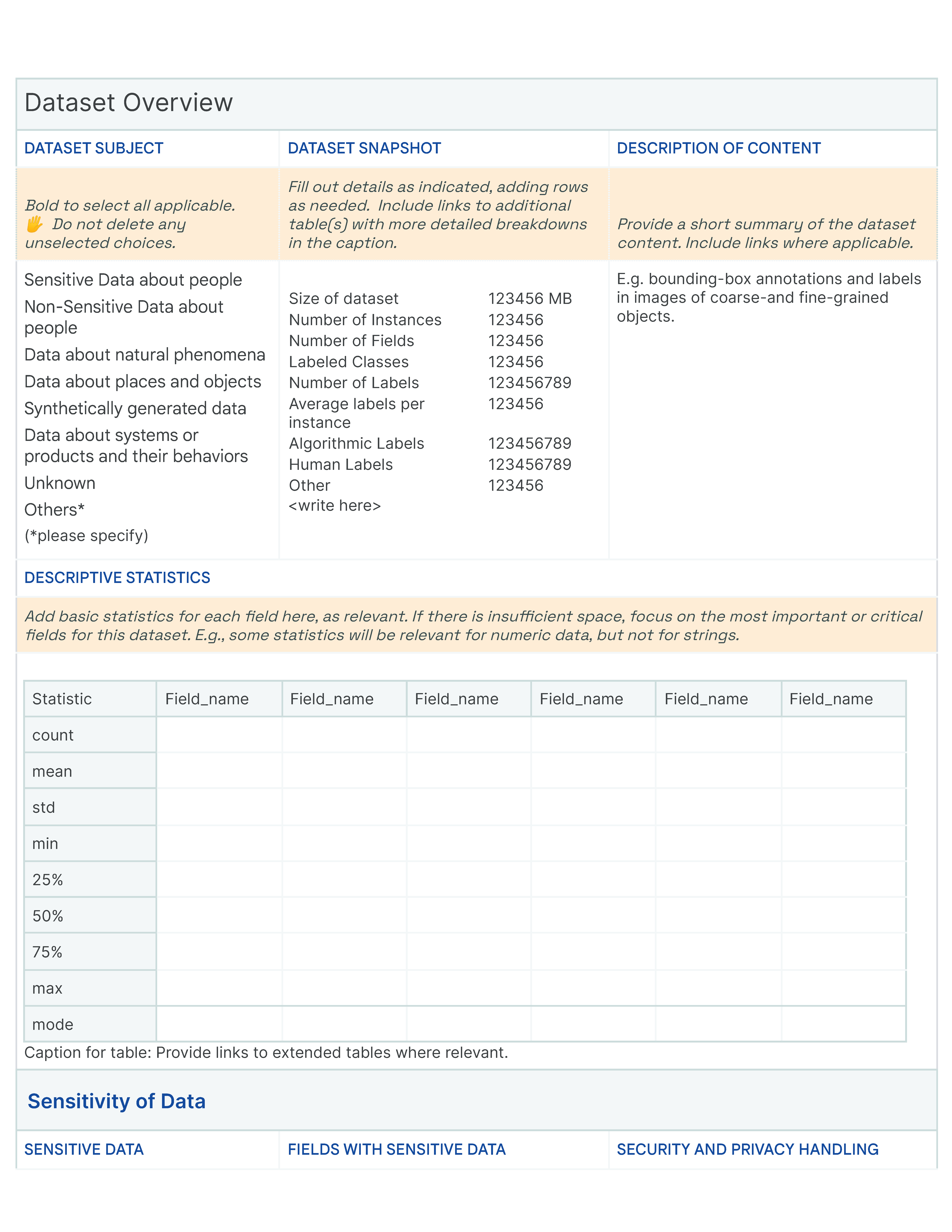}
    \caption{Data Card Template - The \textit{Dataset Overview} section (1/3) of the Data Card was designed as a top-level summary of the dataset that could be included within other transparency artifacts. In those cases, we encourage producers to include a link to a more complete Data Card with other sections. }
    \label{fig:dct-p3}
\end{figure}

\begin{figure}[!htbp]
    \centering
    \includegraphics[width=\textwidth, height=0.9\textheight, frame]{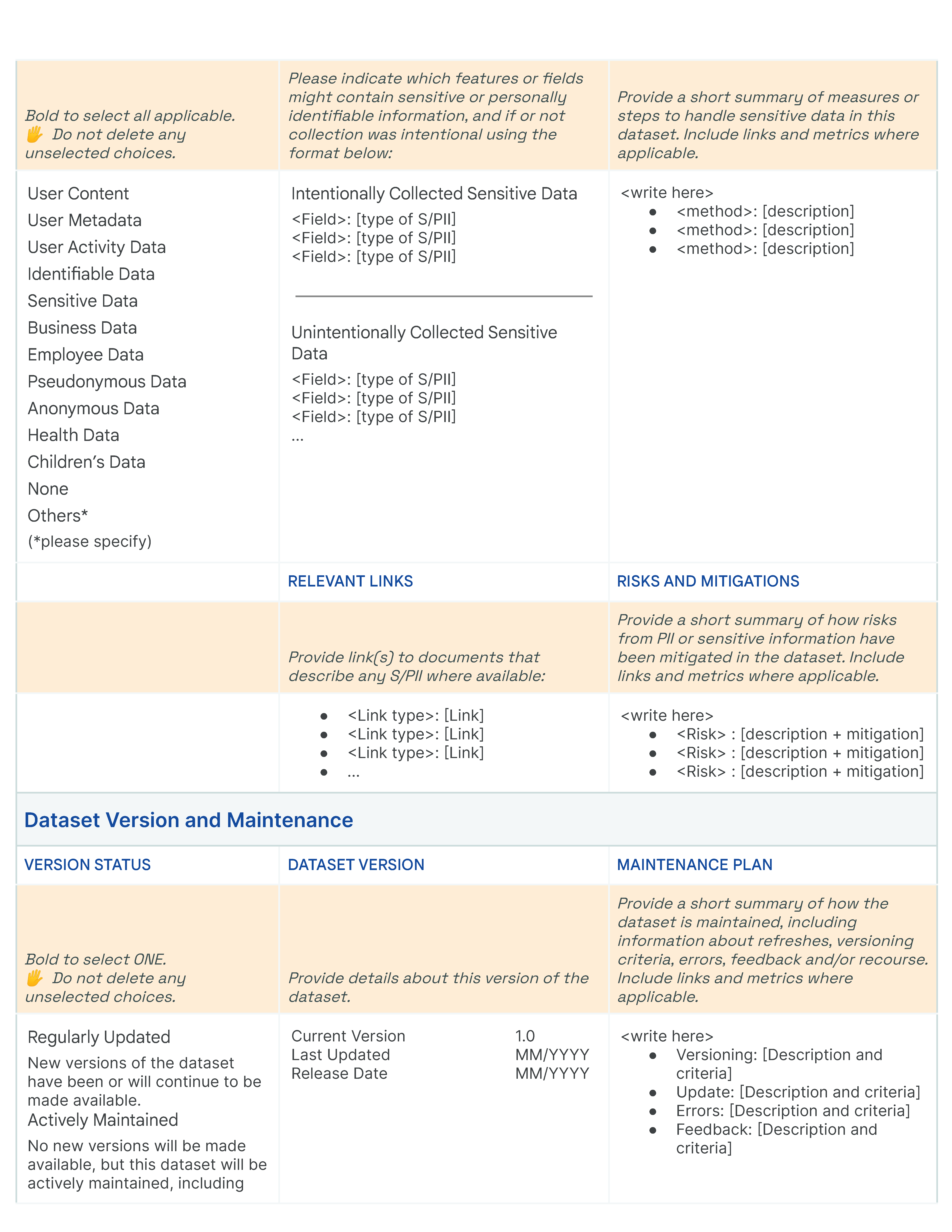}
    \caption{Data Card Template - (Contd., 2/3) The \textit{Sensitivity of Dataset} and \textit{Dataset Version and Maintenance} subsections in the \textit{Dataset Overview} section. The \textit{Sensitivity of Dataset} subsection describes the intentionality, handling, and risks associated with potentially sensitive fields in a dataset. }
    \label{fig:dct-p4}
\end{figure}

\begin{figure}[!htbp]
    \centering
    \includegraphics[width=\textwidth, height=0.9\textheight, frame]{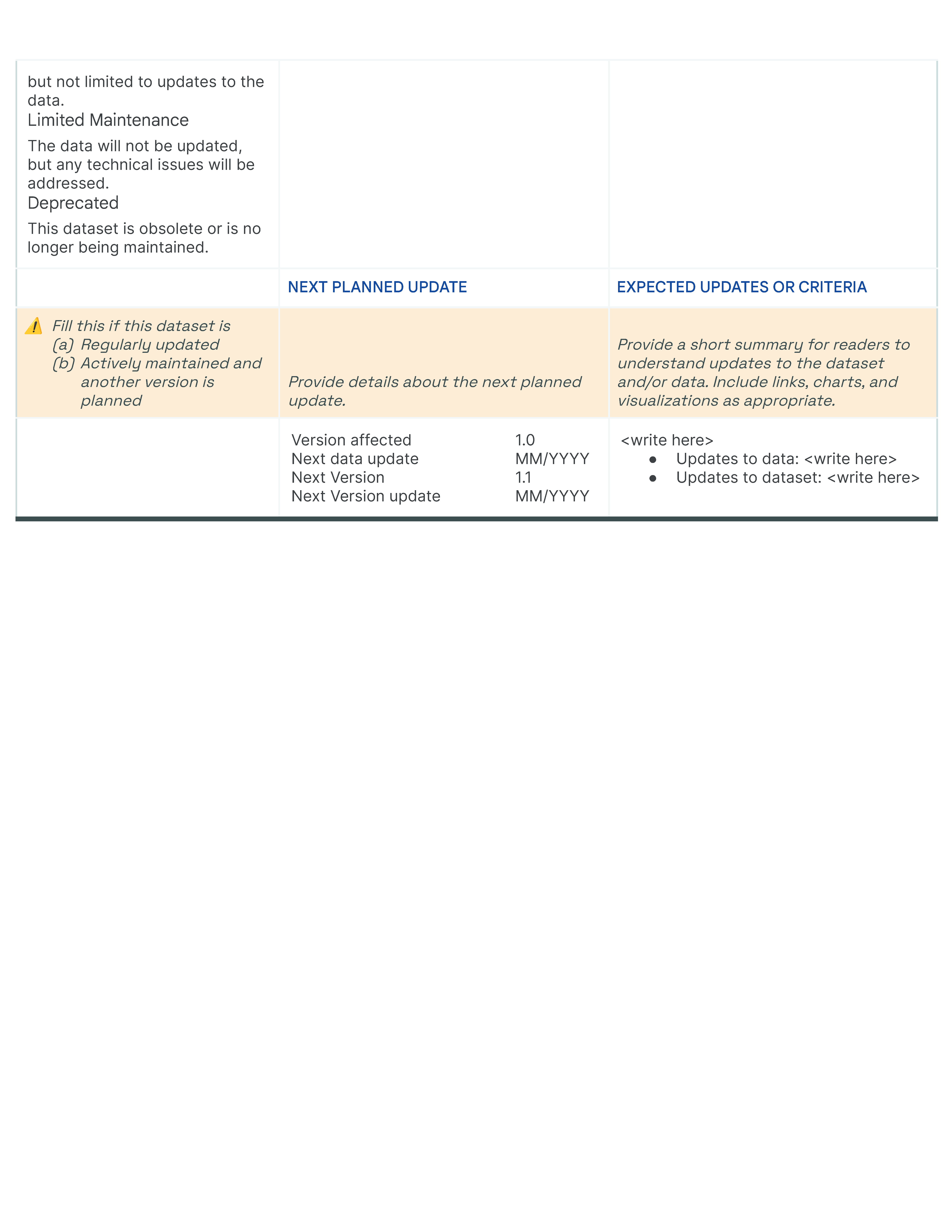}
    \caption{Data Card Template - (Contd., 3/3) The \textit{Dataset Version and Maintenance} subsection in the \textit{Dataset Overview} section.}
    \label{fig:dct-p5}
\end{figure}

\begin{figure}[!htbp]
    \centering
    \includegraphics[width=\textwidth, height=0.9\textheight, frame]{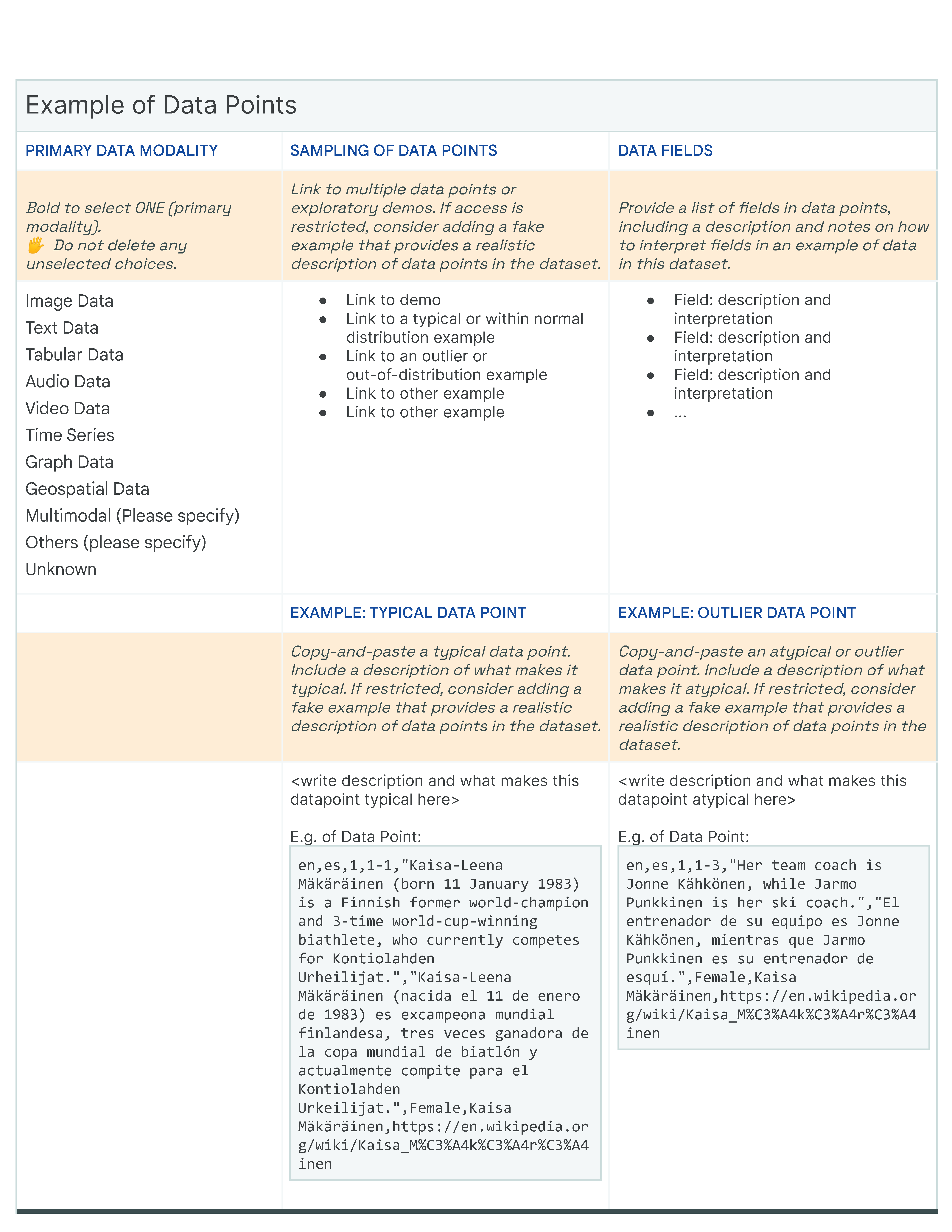}
    \caption{Data Card Template - The \textit{Example of Datapoints} section is designed to help readers interpret and first-hand explore data points in the dataset without needing to download the dataset. This improves both the use of the dataset and usability of the Data Card.}
    \label{fig:dct-p6}
\end{figure}

\begin{figure}[!htbp]
    \centering
    \includegraphics[width=\textwidth, height=0.9\textheight, frame]{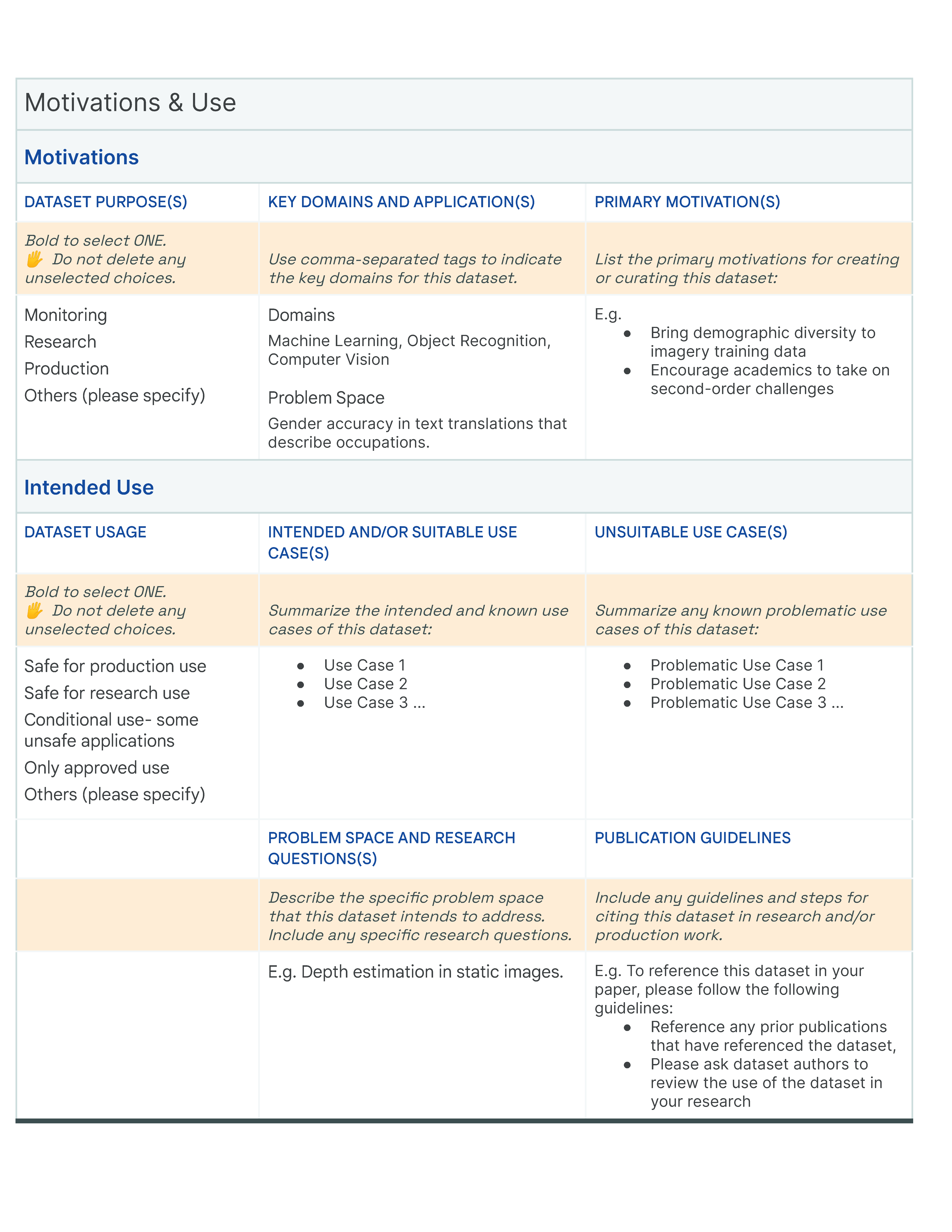}
    \caption{Data Card Template - The \textit{Motivations \& Use} section asks producers to describe their motivations for creating the dataset, as well as the intended uses of the dataset. The \textit{Motivations} subsection sets up the domain of research or application as well as the specific problems the dataset was designed for. We encourage producers to describe \textit{known} suitable and unsuitable use cases for their dataset in the \textit{Intended Use} subsection since it is impossible to list every possible use case of datasets.}
    \label{fig:dct-p7}
\end{figure}

\begin{figure}[!htbp]
    \centering
    \includegraphics[width=\textwidth, height=0.9\textheight, frame]{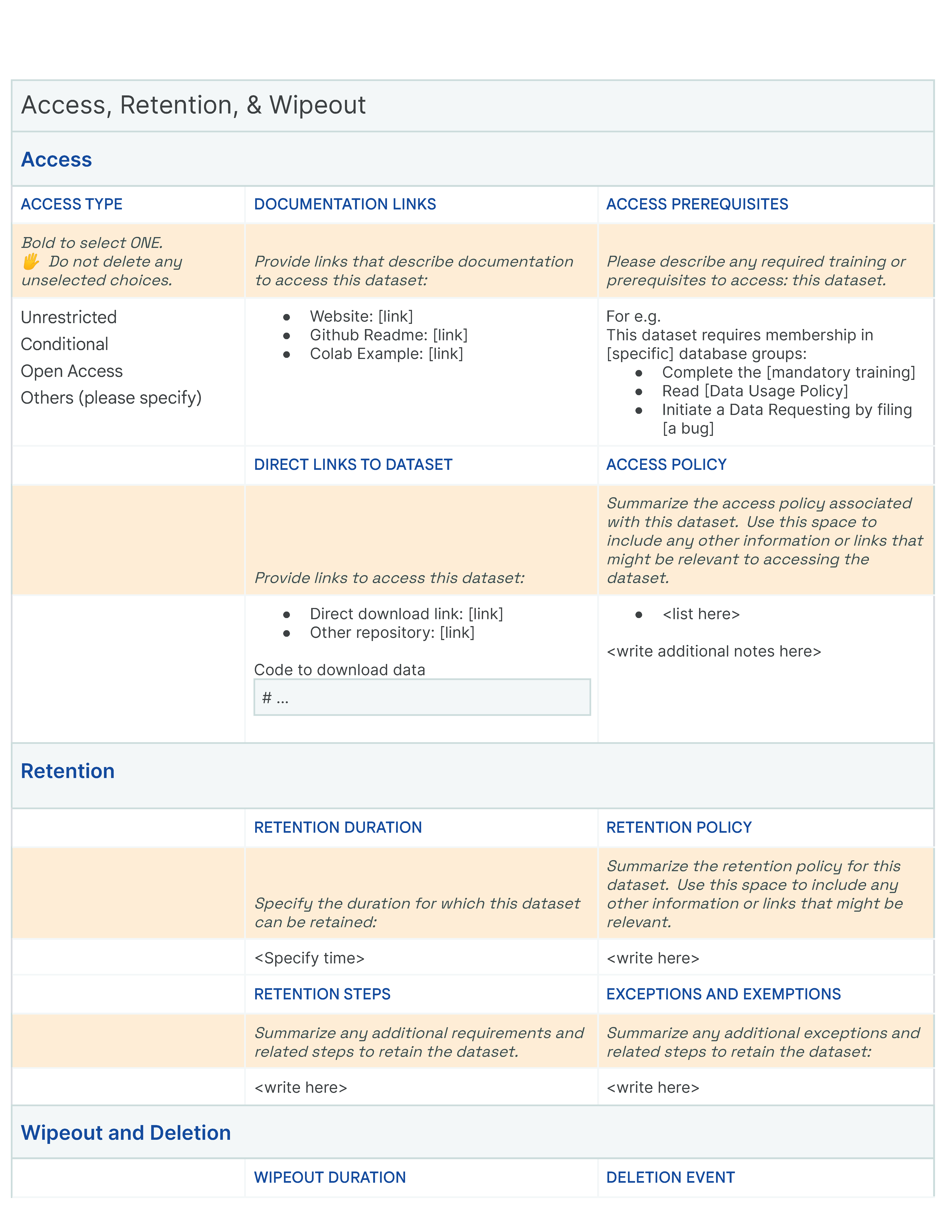}
    \caption{Data Card Template - The \textit{Access, Retention and Wipeout} section (1/2) is decomposed into separate subsections. The \textit{Access} subsection details the storage locations of the dataset, as well as any pre-requisites and policies that govern access to the dataset. This is particularly important for regulated industries. The \textit{Retention} subsection describes the retention duration and summarizes the retention policies and exceptions that are applicable to the dataset. }
    \label{fig:dct-p8}
\end{figure}

\begin{figure}[!htbp]
    \centering
    \includegraphics[width=\textwidth, height=0.9\textheight, frame]{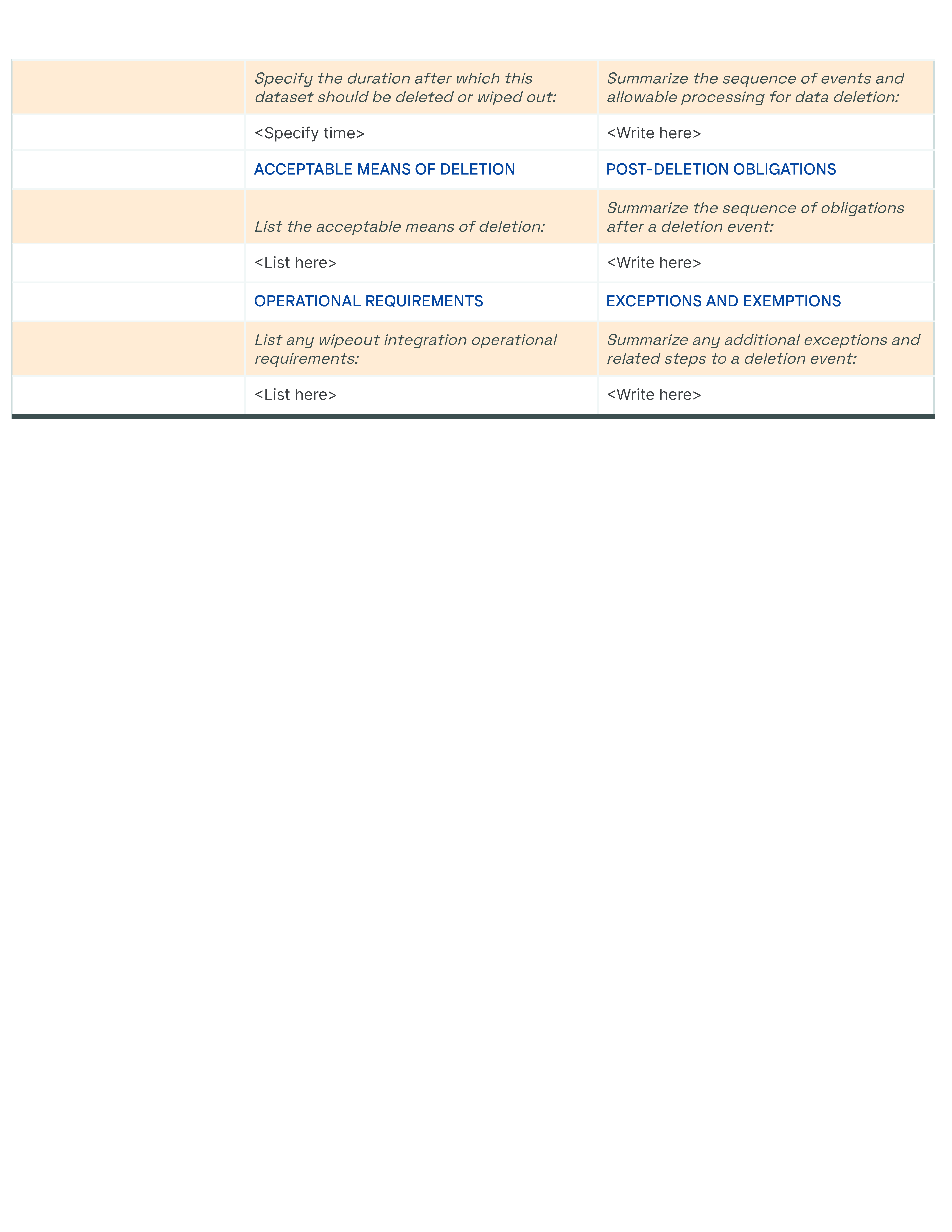}
    \caption{Data Card Template - The \textit{Access, Retention and Wipeout} section (Contd., 2/2) include a subsection on \textit{Wipeout and Deletion} to provide guidance on the most appropriate way to delete a dataset after the retention period has expired. It also asks producers to include information about exceptions and exemptions to wipeout policies.}
    \label{fig:dct-p9}
\end{figure}

\begin{figure}[!htbp]
    \centering
    \includegraphics[width=\textwidth, height=0.9\textheight, frame]{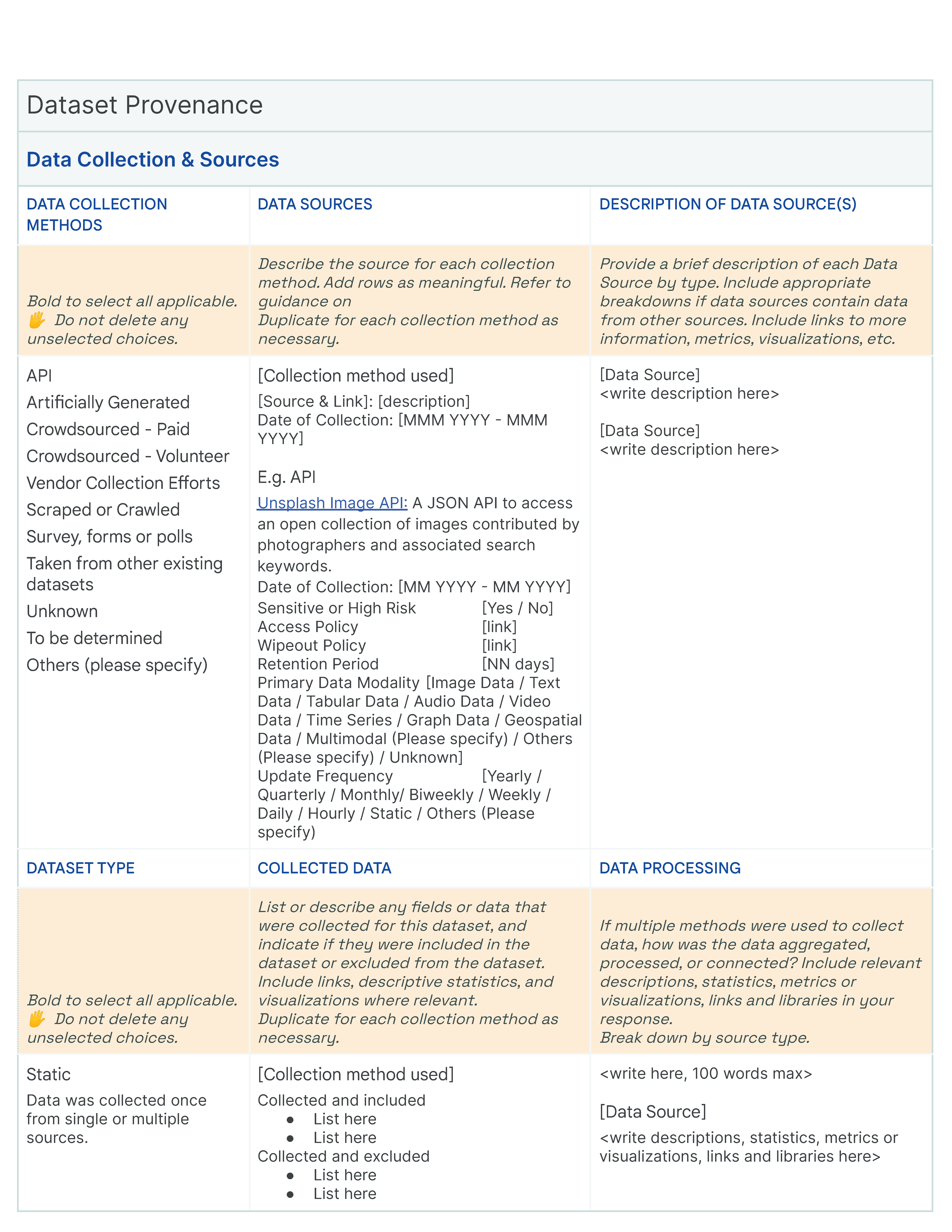}
    \caption{Data Card Template - The \textit{Dataset Provenance} section (1/3) describes the origin of the datasets using subsections. The \textit{Data Collection and Sources} subsection provides an overview that describes several qualitative and procedural attributes of the collection methods and upstream sources of datapoints in the dataset.}
    \label{fig:dct-p10}
\end{figure}

\begin{figure}[!htbp]
    \centering
    \includegraphics[width=\textwidth, height=0.9\textheight, frame]{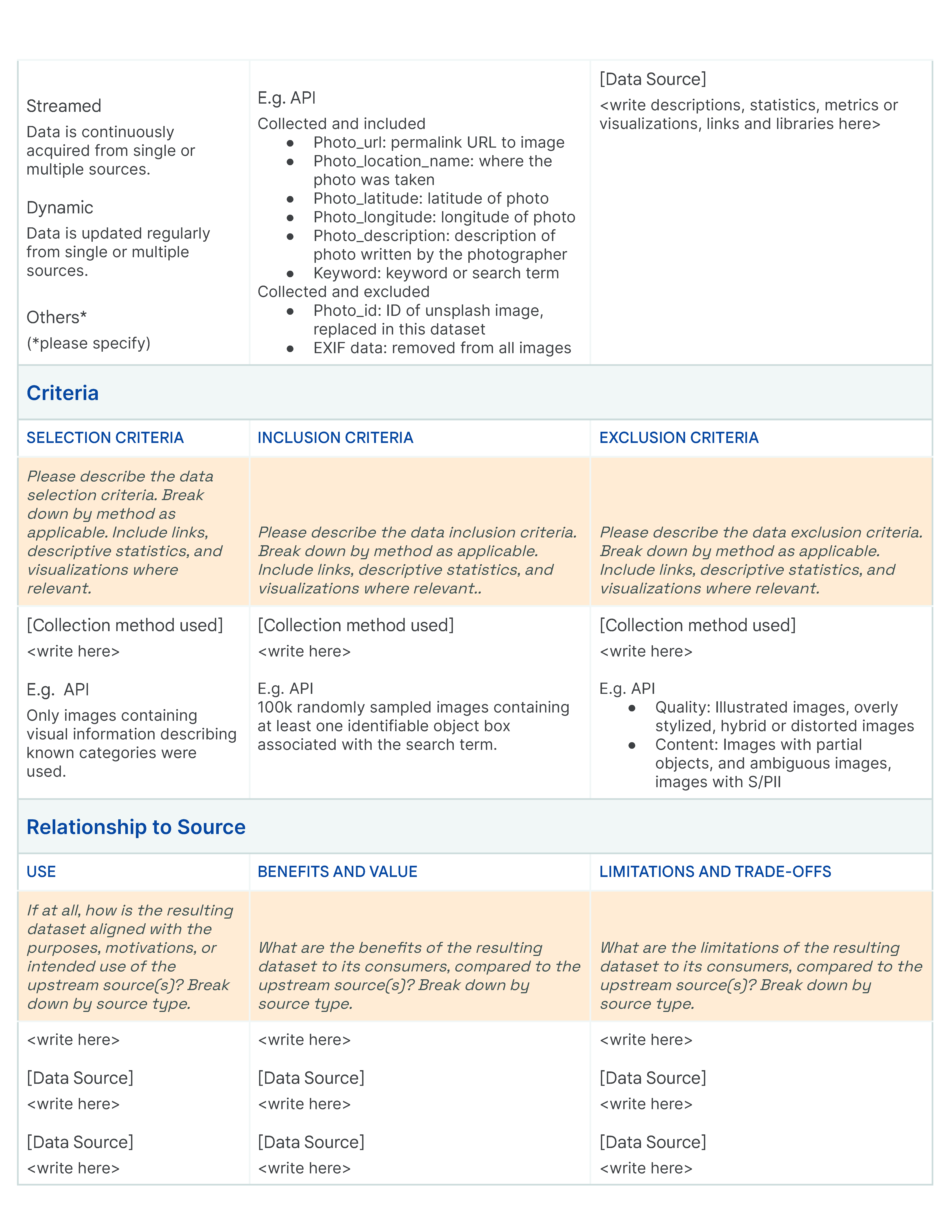}
    \caption{Data Card Template - Within the \textit{Dataset Provenance} section (Contd., 2/3), the \textit{Criteria} subsection elaborates on decisions and parameters pertaining to selection, inclusion, and exclusion of datapoints from the dataset, while the \textit{Relationship to Source} subsection establishes the nature of upstream sources of datapoints in the dataset. Both subsections have been designed to account for multiple collection methods and upstream sources, particularly relevant where datasets have been created through aggregation or joining.}
    \label{fig:dct-p11}
\end{figure}

\begin{figure}[!htbp]
    \centering
    \includegraphics[width=\textwidth, height=0.9\textheight, frame]{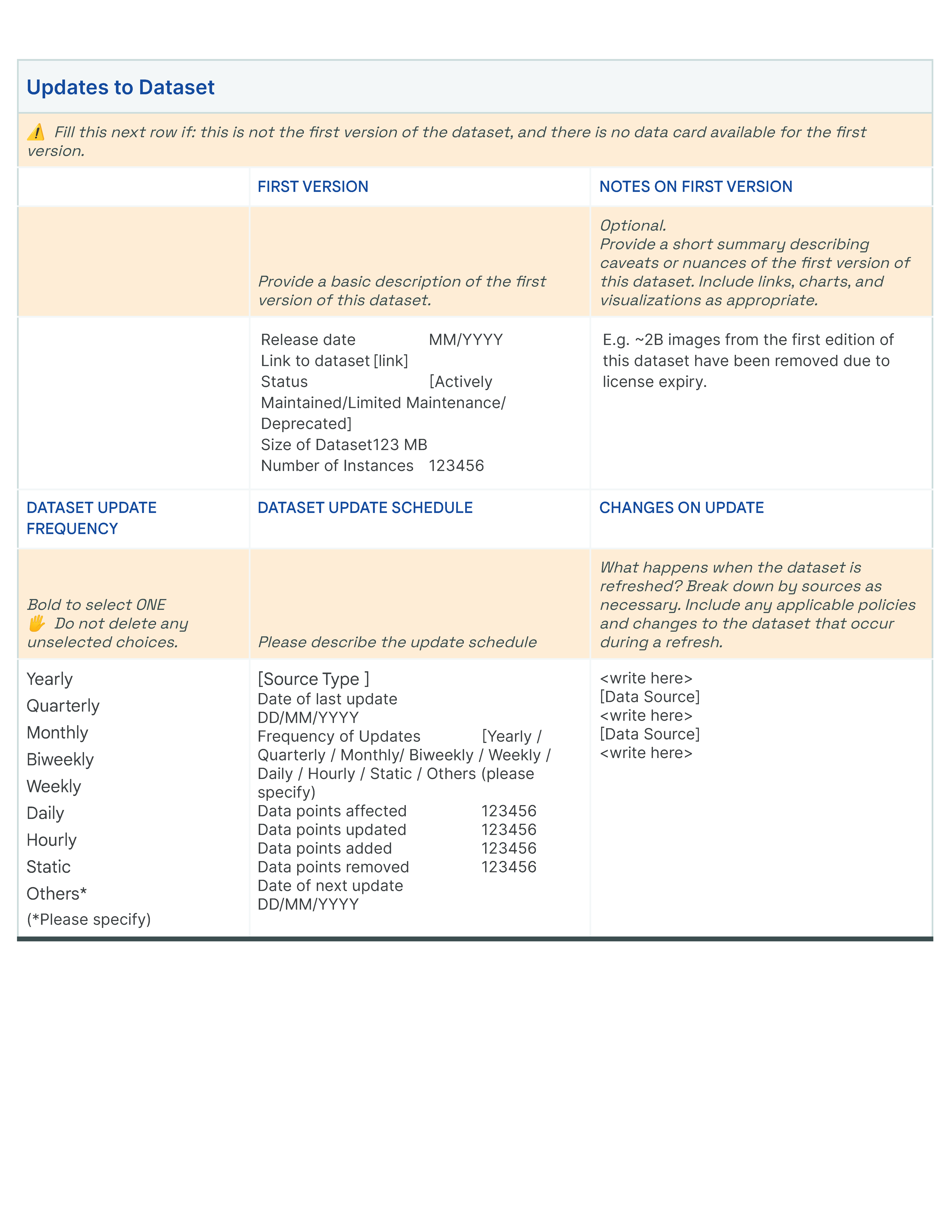}
    \caption{Data Card Template - (Contd., 3/3) In practice we find that producers find it easier to create Data Cards for new dataset or new versions of existing datasets, rather than retroactively creating data cards for previous versions. This decision has been frequently attributed to the loss of knowledge to time. The \textit{Updates to Dataset} subsection is a part of the \textit{Data Provenance} section, and is designed to capture nuances of the most recent updates to the dataset, and plans for future updates to the dataset.}
    \label{fig:dct-p12}
\end{figure}

\begin{figure}[!htbp]
    \centering
    \includegraphics[width=\textwidth, height=0.9\textheight, frame]{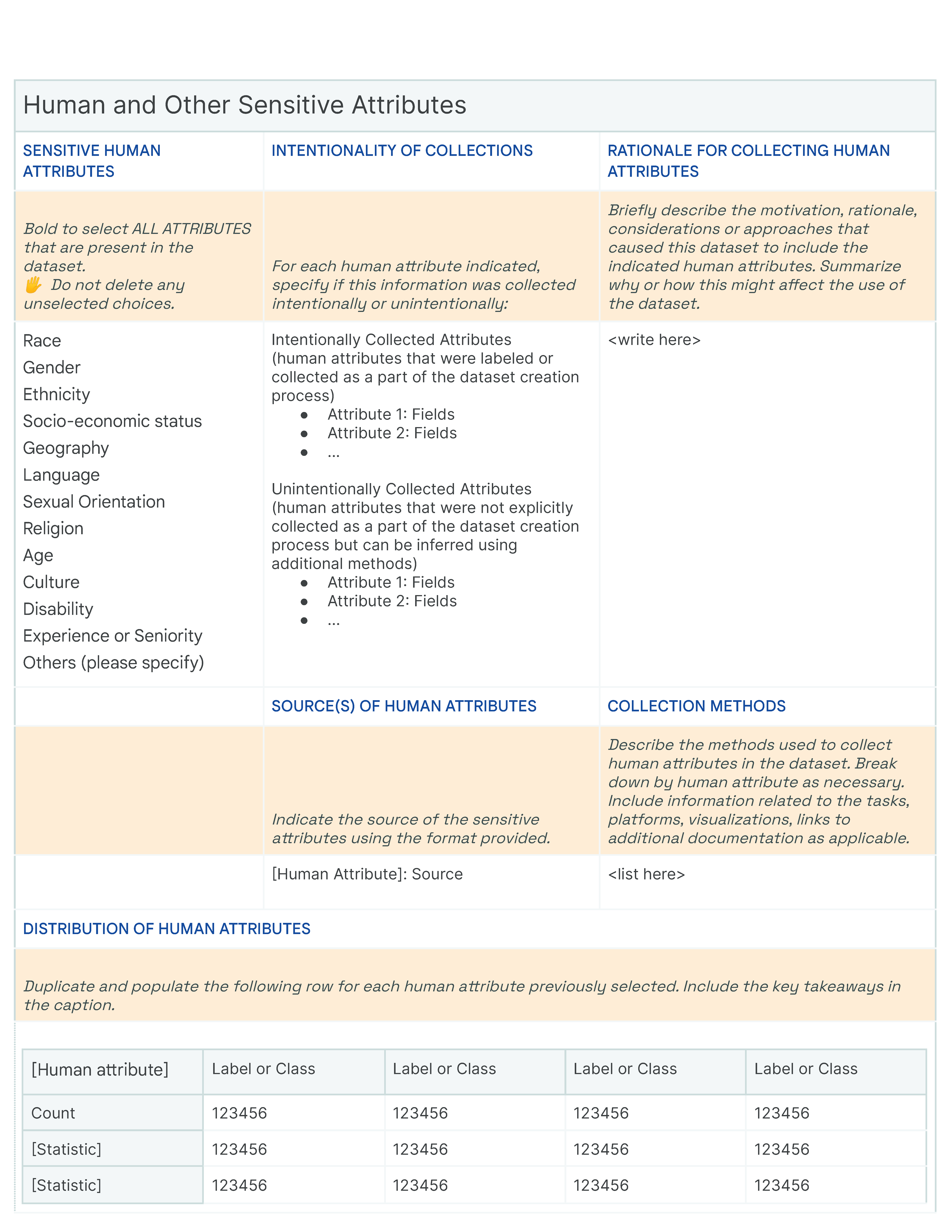}
    \caption{Data Card Template - The \textit{Human and Other Sensitive Attributes} (1/2) is of particular importance to human-centered machine learning applications and fairness analyses. Here, we encourage producers to report the rationales behind decisions to capture or include human attributes as well as various disaggregated statistics and correlations, risks and trade-offs (see Figure \ref{fig:dct-p14}).}
    \label{fig:dct-p13}
\end{figure}

\begin{figure}[!htbp]
    \centering
    \includegraphics[width=\textwidth, height=0.9\textheight, frame]{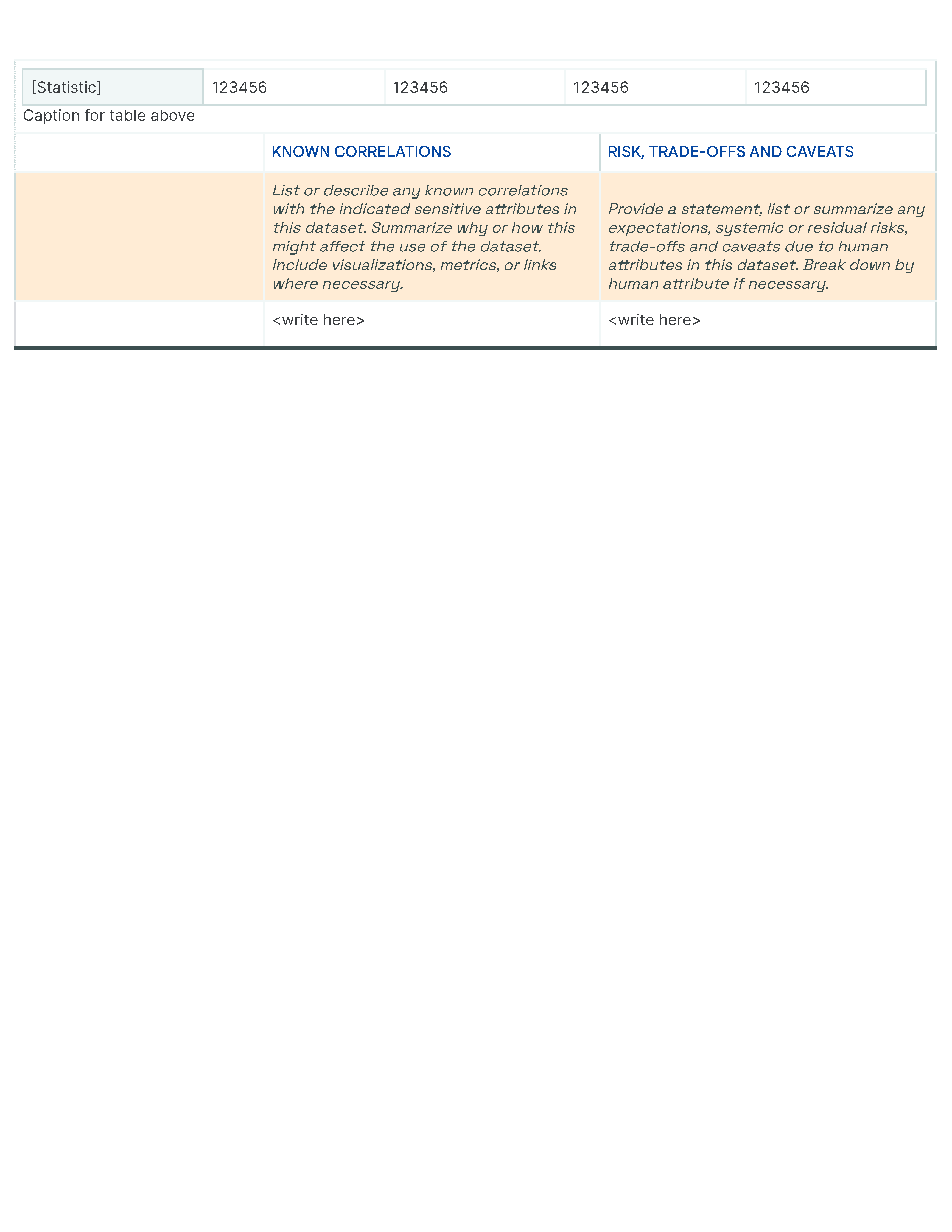}
    \caption{Data Card Template - The \textit{Human and Other Sensitive Attributes} (Contd., 2/2)}
    \label{fig:dct-p14}
\end{figure}

\begin{figure}[!htbp]
    \centering
    \includegraphics[width=\textwidth, height=0.9\textheight, frame]{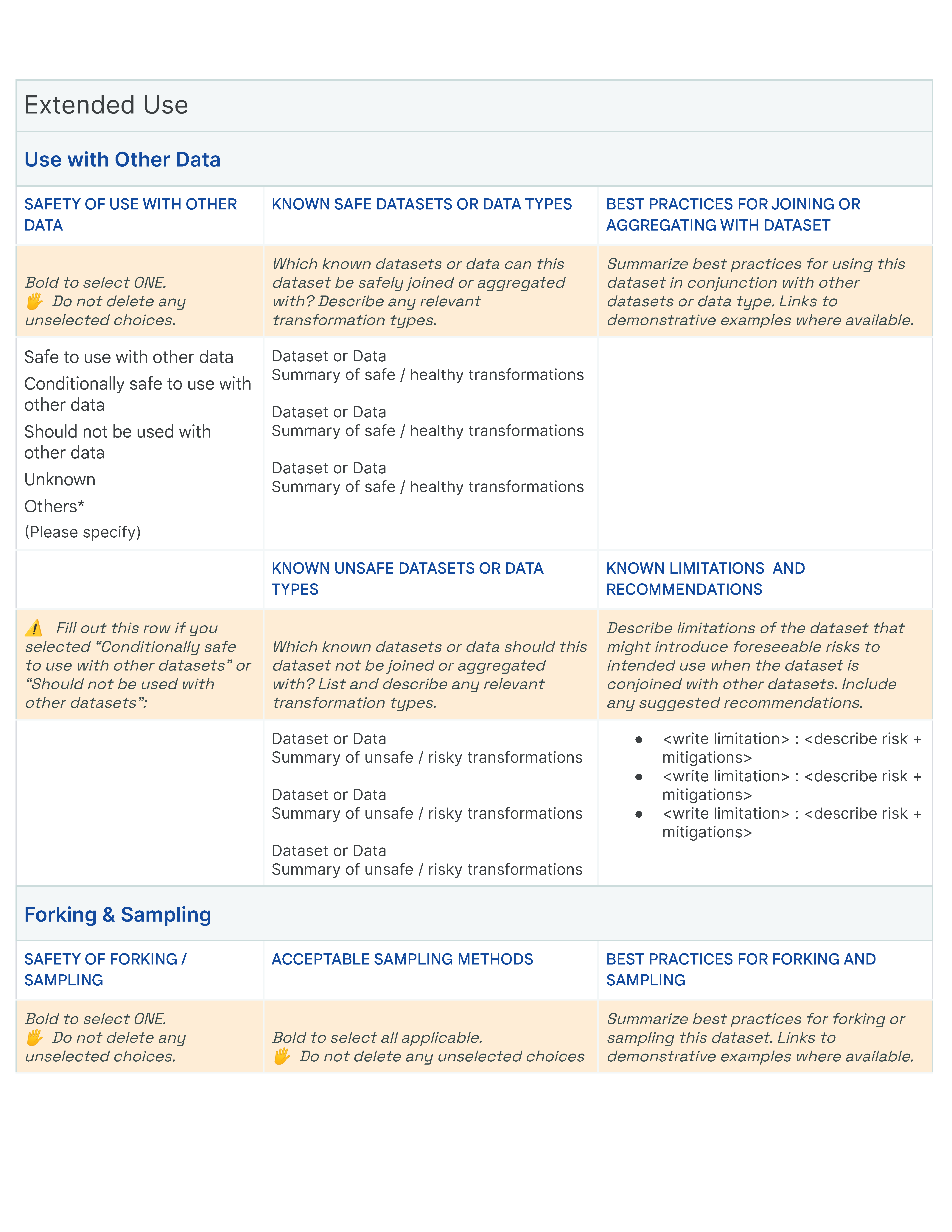}
    \caption{Data Card Template - The \textit{Extended Use} section (1/3) is designed to capture guidance necessary for the responsible use of the dataset, including joining with other datasets or forking and sampling the dataset. }
    \label{fig:dct-p15}
\end{figure}

\begin{figure}[!htbp]
    \centering
    \includegraphics[width=\textwidth, frame]{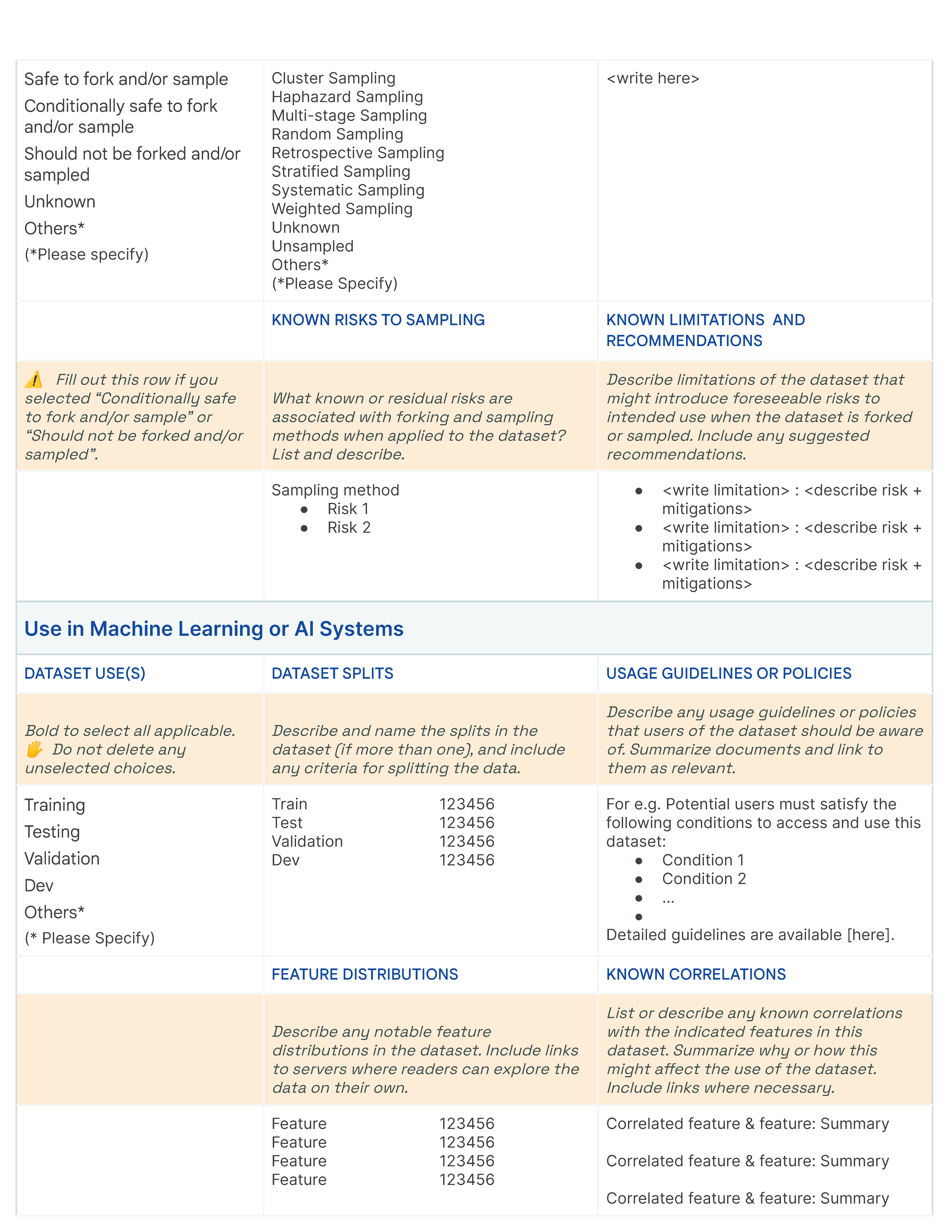}
    \caption{Data Card Template - The \textit{Extended Use} section (Contd., 2/3) captures known or benchmark performance when the dataset is used in various ML applications in the \textit{Use in Machine Learning or AI systems} subsection. }
    \label{fig:dct-p16}
\end{figure}

\begin{figure}[!htbp]
    \centering
    \includegraphics[width=\textwidth, frame]{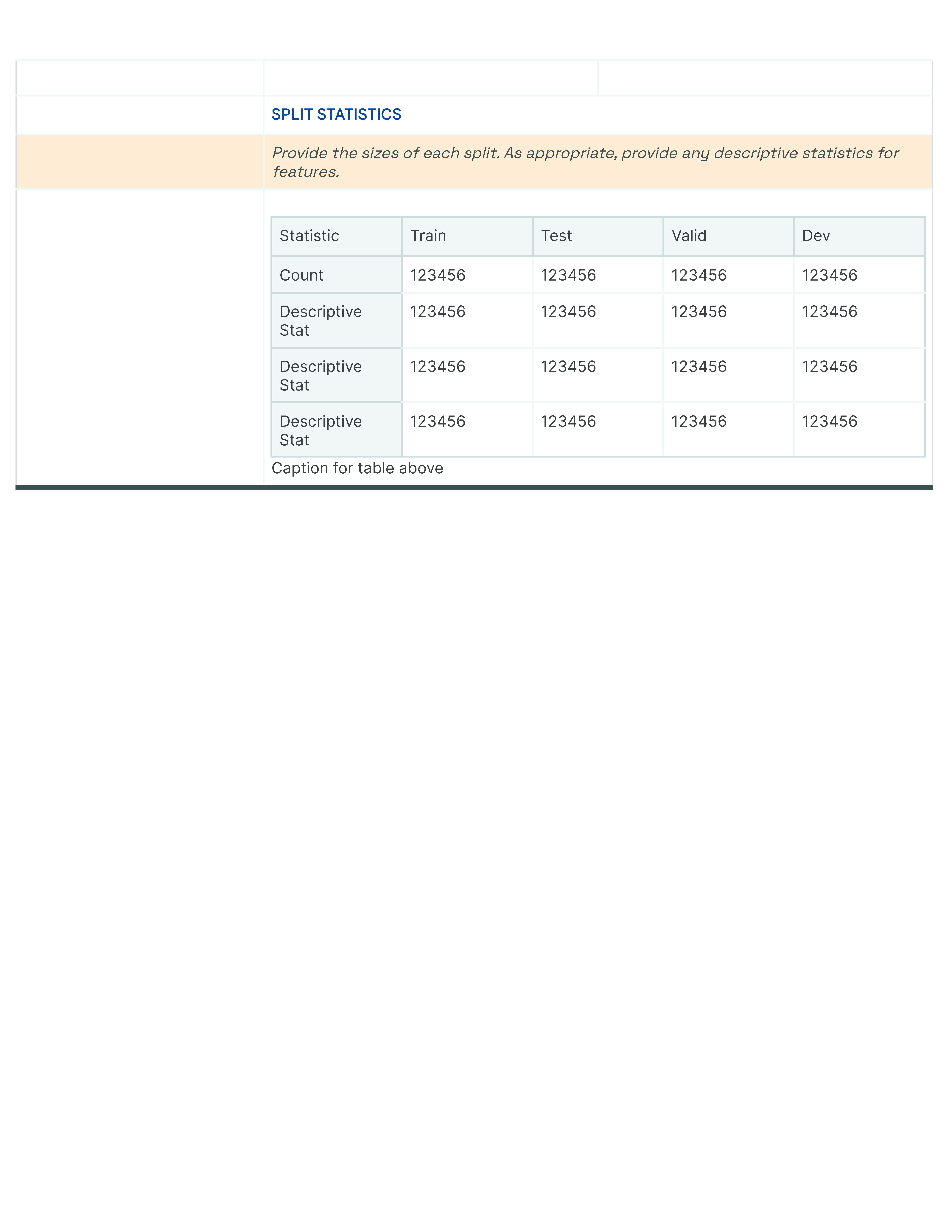}
    \caption{Data Card Template - In the \textit{Use in Machine Learning or AI systems} subsection of the \textit{Extended Use} section (Contd., 3/3), producers are ask to report descriptive statistics for different training and testing splits. For widescale adoption, we encourage the automation of these types of fields for accuracy and rigor.}
    \label{fig:dct-p17}
\end{figure}

\begin{figure}[!htbp]
    \centering
    \includegraphics[width=\textwidth, height=0.9\textheight, frame]{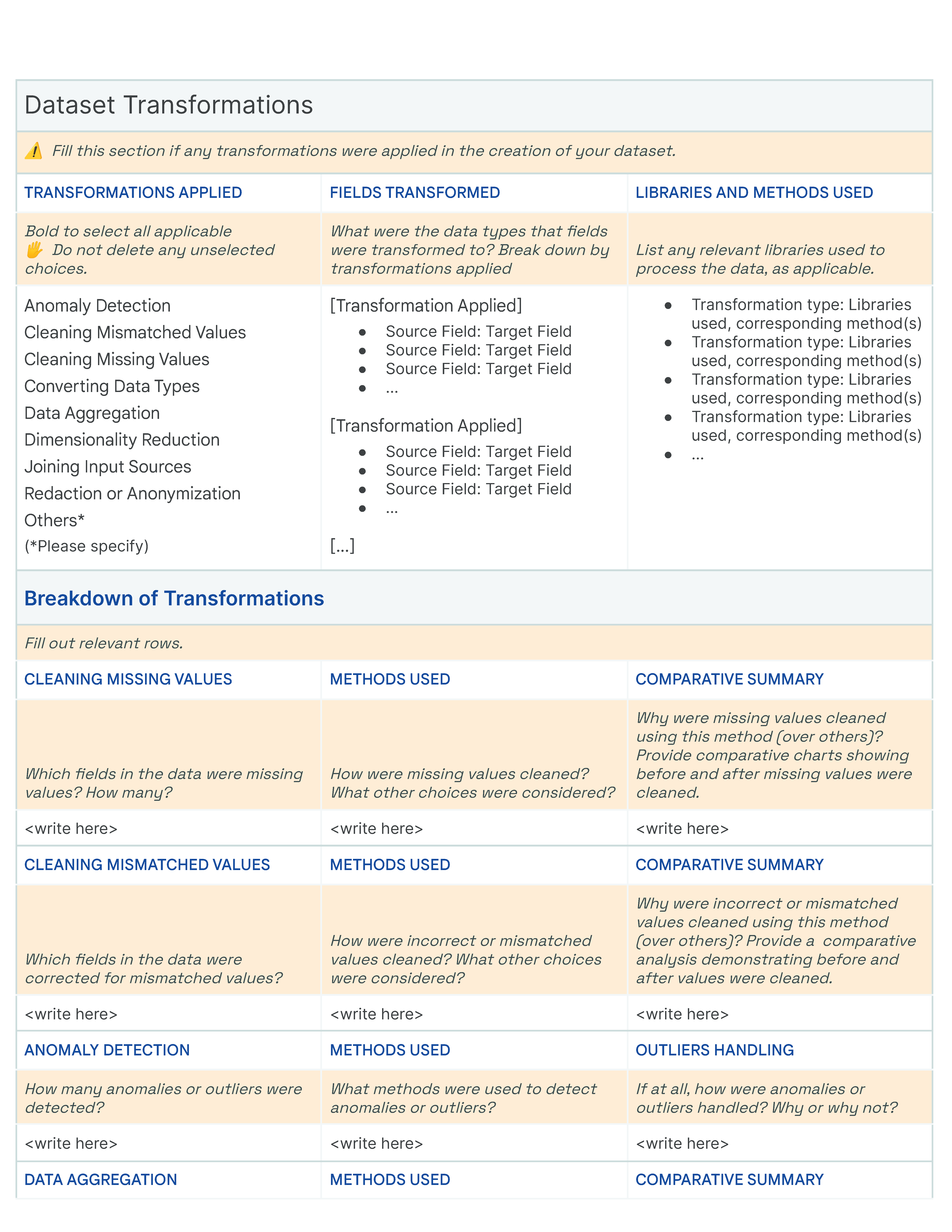}
    \caption{Data Card Template - The \textit{Transformations} section is used to describe the processes by which raw data is transformed into usable formats. Here, we first ask producers to provide a aggregate of the transformations, following which a more detailed breakdowns are collected.}
    \label{fig:dct-p18}
\end{figure}

\begin{figure}[!htbp]
    \centering
    \includegraphics[width=\textwidth, height=0.9\textheight, frame]{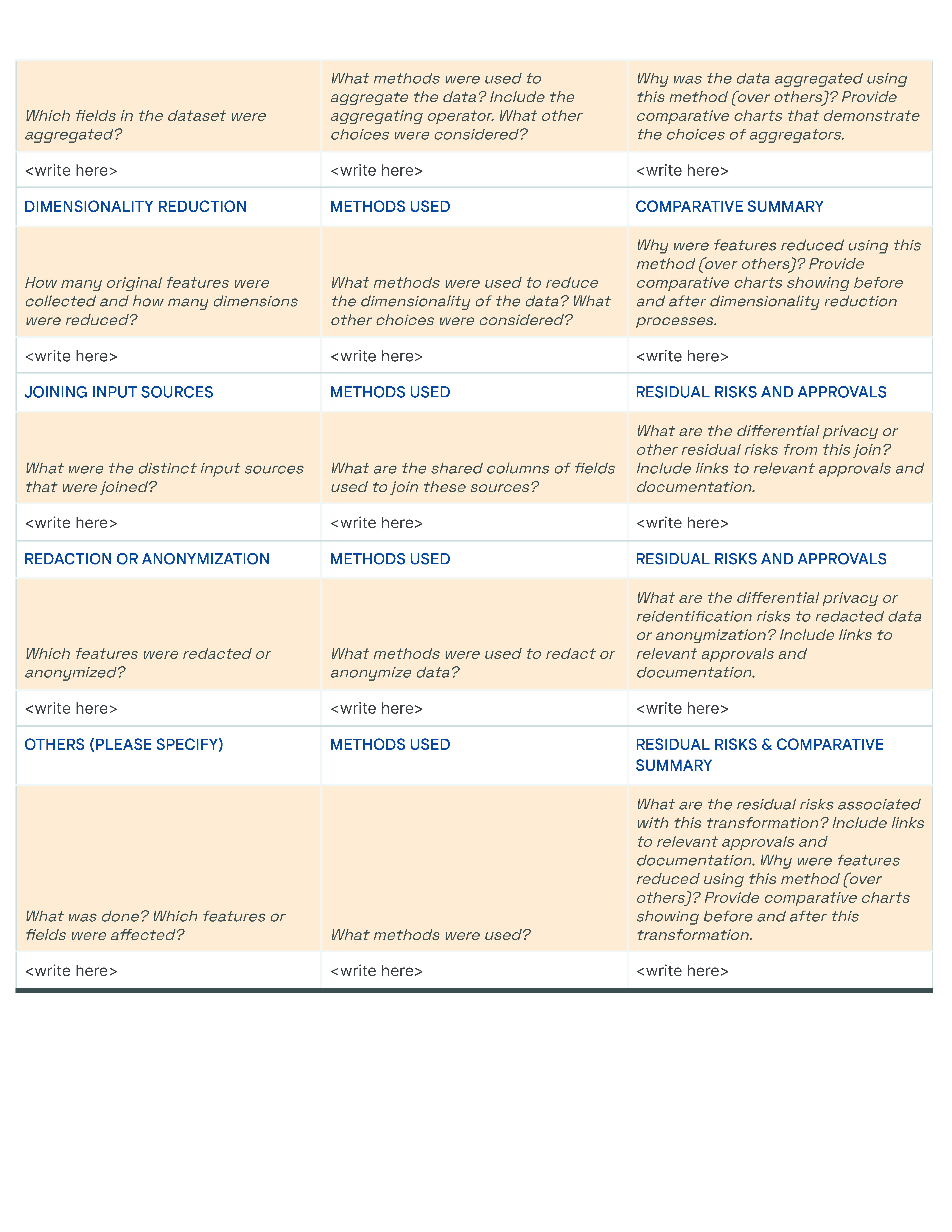}
    \caption{Data Card Template - The \textit{Transformations} section (Contd.). Producers are asked to include information about specific transformation applied to datasets that could potentially introduce residual or system-level risks and require oversight.}
    \label{fig:dct-p19}
\end{figure}

\begin{figure}[!htbp]
    \centering
    \includegraphics[width=\textwidth, height=0.9\textheight, frame]{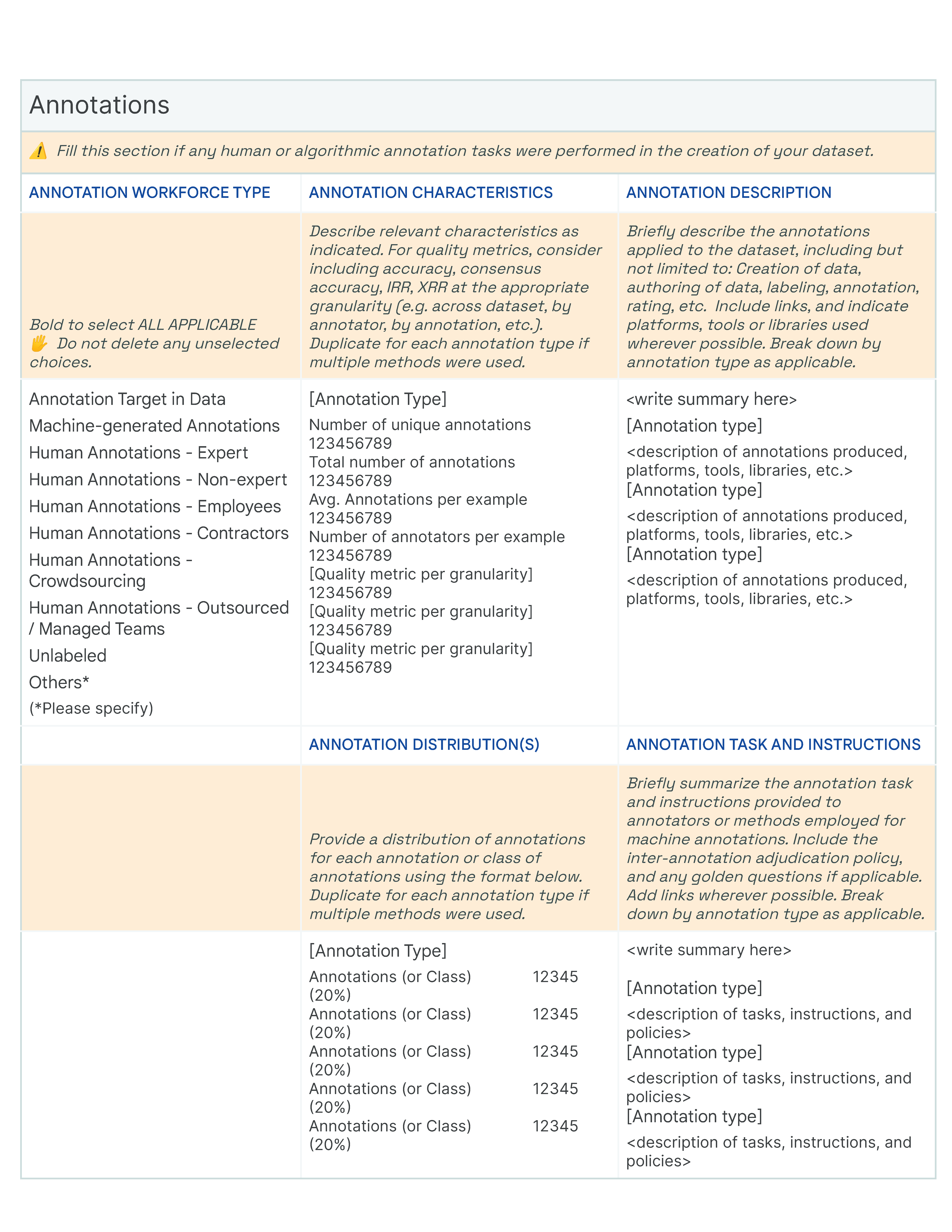}
    \caption{Data Card Template - The \textit{Annotations} section captures a variety of annotation types, including quantiative characteristics, qualitative descriptions, resulting distributions, and task or instruction summaries that affect outcomes.}
    \label{fig:dct-p20}
\end{figure}

\begin{figure}[!htbp]
    \centering
    \includegraphics[width=\textwidth, height=0.9\textheight, frame]{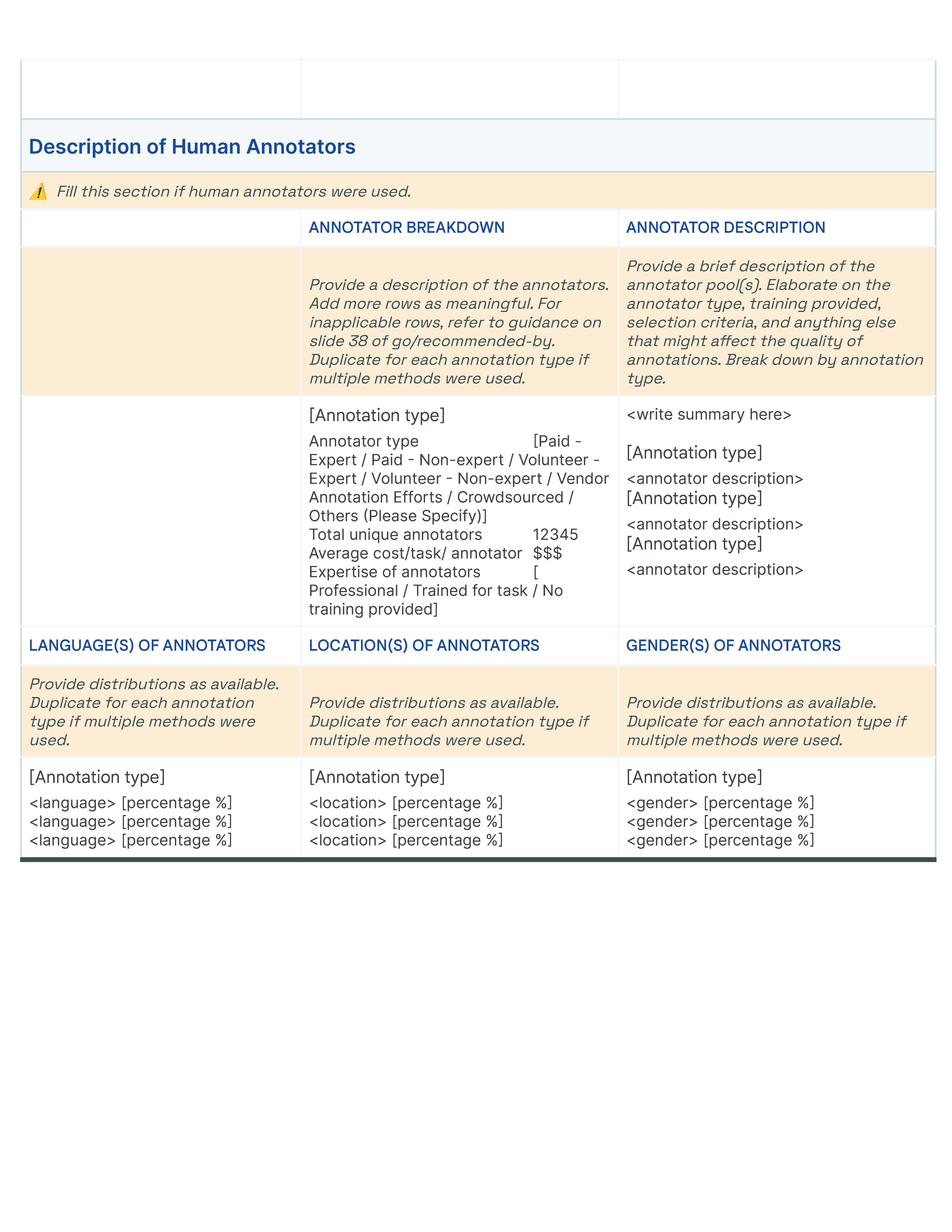}
    \caption{Data Card Template - Important to human computation datasets, this sub-section of the \textit{Annotations} section captures attributes where human annotators were employed.}
    \label{fig:dct-p21}
\end{figure}

\begin{figure}[!htbp]
    \centering
    \includegraphics[width=\textwidth, height=0.9\textheight, frame]{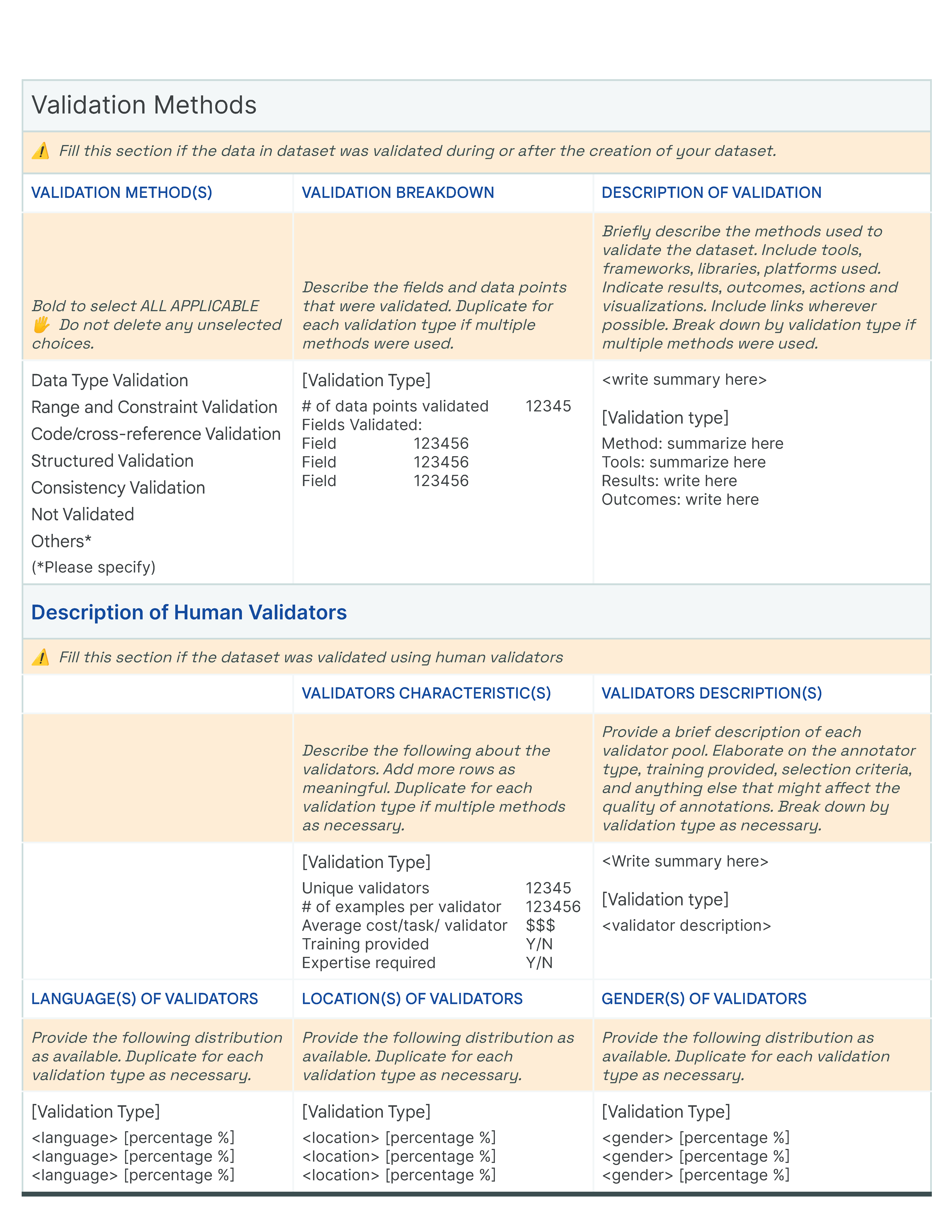}
    \caption{Data Card Template - Producers are expected to complete the \textit{Validation Methods} section if a part or the entirety of the dataset was validated. This section also details attributes of human validators.  }
    \label{fig:dct-p22}
\end{figure}

\begin{figure}[!htbp]
    \centering
    \includegraphics[width=\textwidth, height=0.9\textheight, frame]{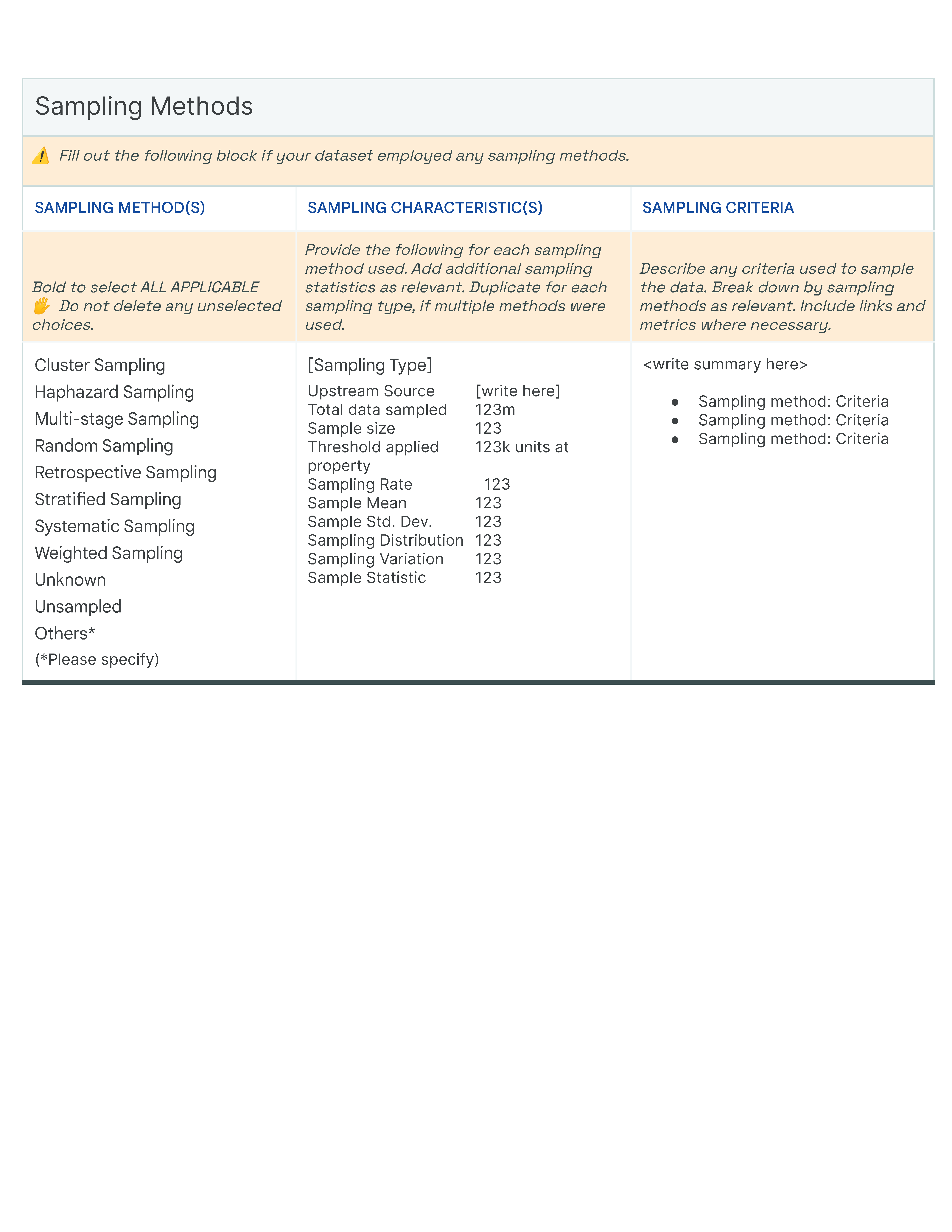}
    \caption{Data Card Template -  The \textit{Sampling Methods} section captures both quantitative metrics pertinent and qualitative summaries pertinent to sampling that may have been used in the creation of the dataset. Since not all datasets may be sampled, this section is considered conditional.}
    \label{fig:dct-p23}
\end{figure}

\begin{figure}[!htbp]
    \centering
    \includegraphics[width=\textwidth, frame]{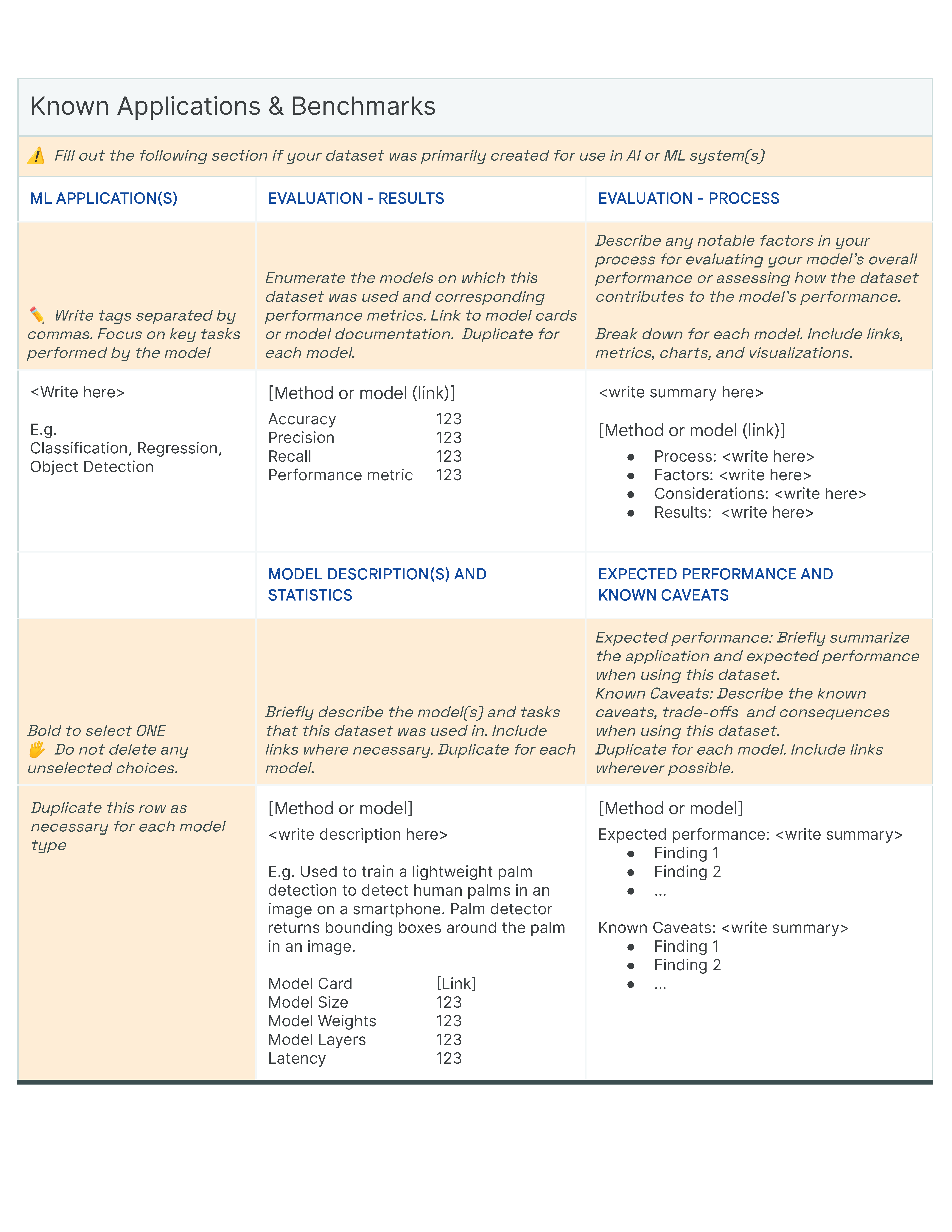}
    \caption{Data Card Template - The \textit{Known Applications \& Benchmarks} section is designed to capture documentation pertaining to the use of the dataset to train or test models, for example, those that are publicly available. Producers are asked to provide a brief description of the model(s), the evaluation processes, expected performance and any known caveats that agents should be aware of.}
    \label{fig:dct-p24}
\end{figure}

\begin{figure}[!htbp]
    \centering
    \includegraphics[width=\textwidth, height=0.9\textheight, frame]{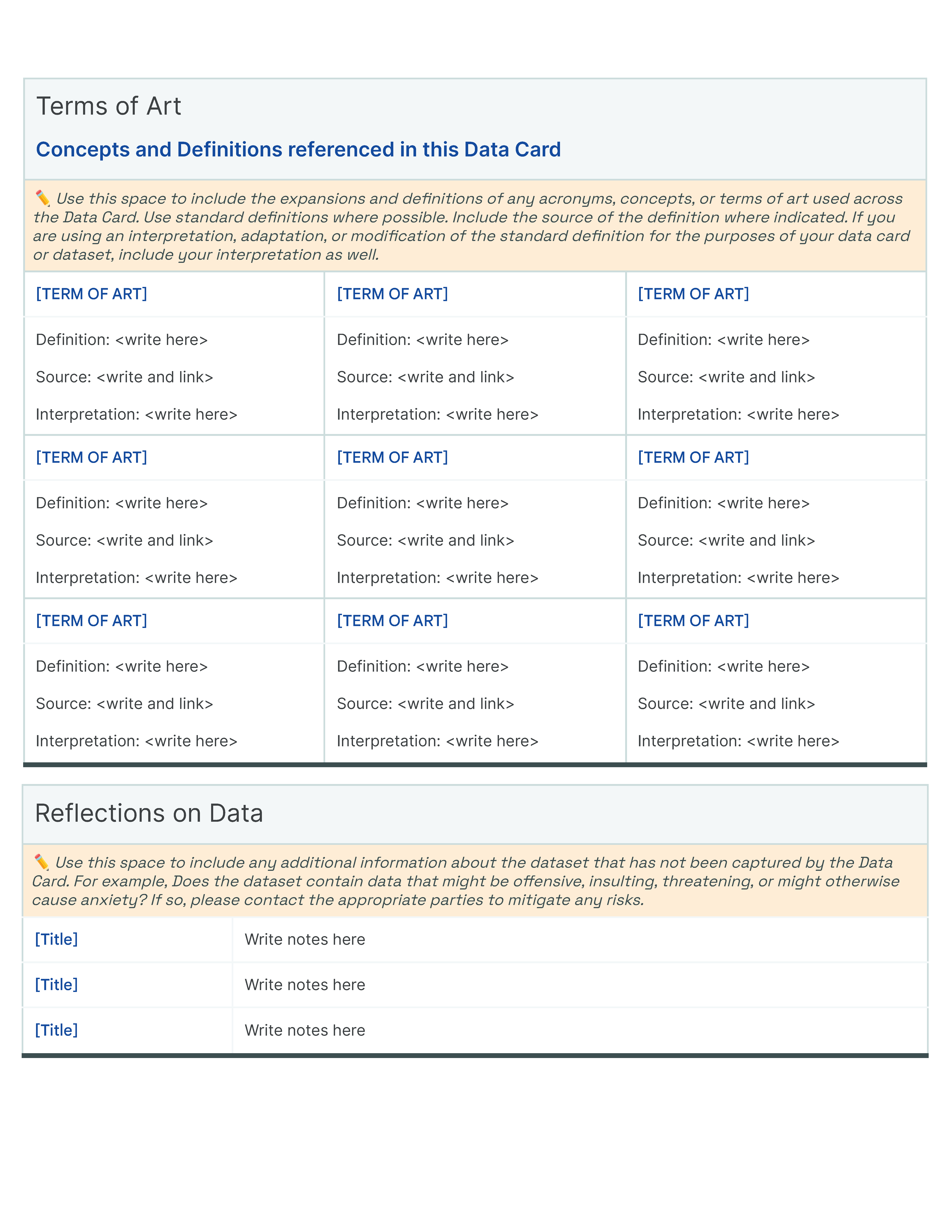}
    \caption{Data Card Template - The \textit{Terms of Art} section introduces technical terms, domain-specific concepts, and acronyms that are used across the Data Card. Here, we ask producers to include any modifications or adaptations to terms to assist with interpretation in the context of the dataset. Tge \textit{Reflections on Data} section is intended to be a free-form space for producers to add information not captured by the template.}
    \label{fig:dct-p25}
\end{figure}

\end{document}